\documentclass[twocolumn,trackchanges]{aastex631}
\usepackage{wrapfig}  
\usepackage{appendix}

\begin{document}

\title{The Age--Thickness Relation as a Tracer of the Merger History of Disk Galaxies}
\author{Lekshmi Thulasidharan}
\affiliation{Department of Physics, University of Wisconsin-Madison, USA\\}
\author{Elena D'Onghia}
\affiliation{Department of Physics, University of Wisconsin-Madison, USA\\}
\affiliation{Department of Astronomy, University of Wisconsin-Madison, USA\\}
\author{Robert Benjamin}
\affiliation{Department of Astronomy, University of Wisconsin-Madison, USA\\}
\author{Ronald Drimmel}
\affiliation{Osservatorio Astrofisico di Torino, Istituto Nazionale di Astrofisica (INAF), I-10025 Pino Torinese, Italy\\}
\author{Eloisa Poggio}
\affiliation{Osservatorio Astrofisico di Torino, Istituto Nazionale di Astrofisica (INAF), I-10025 Pino Torinese, Italy\\}

\begin{abstract}
In the hierarchical framework of galaxy formation, disk galaxies are shaped by a sequence of mergers and interactions, yet reconstructing this history from observations remains challenging. We show that the merger history of a galaxy leaves measurable imprints in the vertical structure of its stellar disk. Using Milky Way analogues from the TNG50 simulations, we demonstrate that the Age--thickness relation, quantified by the dispersion of vertical stellar positions ($\Delta_z$), encodes both the dynamical heating of pre-existing stars and the birth conditions of stars formed during perturbed phases. Major mergers produce pronounced, step-like features in the Age--$\Delta_z$ relation, reflecting strong disk heating and subsequent re-formation of a thin disk, while flyby interactions generate weaker, localized enhancements associated primarily with disturbed star formation. We show that this diagnostic is robust across different locations within the disk and largely insensitive to fractional distance uncertainties lower than 20\%, though its temporal resolution is limited by uncertainties in stellar ages. Because the Age--$\Delta_z$ relation relies only on stellar positions and ages, it provides an observationally accessible alternative to traditional kinematic diagnostics. With current and upcoming surveys mapping the Milky Way with unprecedented precision, this framework offers a new avenue for reconstructing the merger history of our Galaxy and probing the dynamical evolution of disk galaxies across cosmic time.

\end{abstract}
\section{Introduction} \label{sec:intro}

In the hierarchical theory of galaxy formation, the Milky Way is formed by the merger and accretion of numerous smaller satellite dwarf galaxies over time \citep{white1, white,kaufmann,helmi1}. 
However, the precise timing, number, and mass ratios of the Milky Way’s past merger events remain uncertain.
Central to this challenge is discerning observable imprints of major merger events and gravitational interactions with satellite dwarf galaxies throughout cosmic history. Recent advancements, notably through Gaia \citep{gaia,gaia2023} and SDSS-IV\citep{apogee}, focused on Milky Way stars and have revealed compelling evidence of the relics of a major merger with an ancient galaxy named Gaia-Sausage-Enceladus (GSE) \citep{helmi, bel}. The time of this merger remains uncertain, spanning from 8 to 12 billion years ago  \citep{gilmore,navarro,haywood2018,gallart,lancaster2019,Fattahi_2019,Vincenzo_2019,naidu2021,bel1,das2020,naidu2021,mont,Belokurov_2022,ciuca}.

The conventional method for piecing together the Milky Way's merger history entails integrating stellar kinematics with their metallicities. Extensive data from astrometric sky surveys like Gaia and spectroscopic surveys including APOGEE \citep{apogee}, LAMOST \citep{lamost2012}, GALAH \citep{galah} and others have made the Milky Way an ideal setting for testing theories of galaxy formation and evolution. These past mergers may have left behind trails of stellar streams in the halo of the Milky Way \citep{ibata,helmi1999,majewski_2003,helmi_2006,klement_2009,smith_2009,newberg2009,nissen_2010,weiss_2018}; however, identifying these remnants is challenging in configuration space due to the dynamical mixing with other stars over time. Studies suggest that traces of these events should still be observable in kinematic spaces such as the Energy-Angular momentum space \citep{helmi2000,knebe,Brown_2005,Helmi_2005,Font_2006,Choi_2007,Morrison_2009,gomez2010, Fiorentin_2015,Malhan_2022}. Subsequent examinations using various N-body simulations have shown that detecting such streams kinematically is inefficient due to the difficulty in differentiating in-situ stars from accreted stars. Moreover, satellite energy and angular momentum are not conserved during these interactions, leading to several independent overdensities in kinematic-related spaces and complicating the association with specific merger events \citep{Jean_Baptiste_2017, Pagnini_2023, koppelman2020, Amarante_2022, Khoperskov_2023, Khoperskov_2023b}. High-quality spectroscopic data from various sky surveys has enabled successive studies to combine chemical abundances with kinematics to trace past merger events. One can infer separate instances of accretion by identifying sets of stars with distinct kinematic properties and metallicity distribution functions \citep{Myeong_2019,naidu2021,Ruiz_Lara_2022,Chandra_2023}. However, \cite{mori} suggests a potential bias in this approach, possibly misinterpreting a single accretion event as multiple ones and underestimating the mass of the progenitors. Recent work has further quantified this information loss, showing that the recoverability of merger properties from present-day chemodynamical observables decreases with time as debris becomes dynamically mixed \citep{Necib2026}. While these approaches rely primarily on stellar kinematics and chemical abundances, structural diagnostics of the disk itself remain comparatively underexplored. This highlights the need for complementary approaches that rely on independent observables to reconstruct our Galaxy's accretion history.

It has been known since the early 1990s that satellite interactions can significantly heat the stars in a galaxy's disk, leading to thickening of the disk in the vertical direction \citep{toth, quinn, donghia2010}. 
This suggests that the vertical structure of stellar populations as a function of age provides a fossil record of past perturbations.
In this study, we demonstrate that the merger history of a Milky Way-like galaxy is imprinted in the vertical thickness of its stellar disk. Our analysis focuses on an increase in the vertical thickness of the stellar disk as a key indicator of merger events, drawing support from patterns observed in the TNG50 simulations \citep{nels1}. This idea is analogous to the well-known age--velocity dispersion relation \citep[e.g.,][]{gustaf1946,nordstrom2004,sun2025}, but instead uses structural information encoded in stellar positions.

This paper is structured as follows. In Section \ref{sec:tng50}, we describe the simulations and present the analysis of the thickness of the stellar disk with respect to age. In Section \ref{sec:robustness}, we discuss the robustness of this diagnostic to observational constraints when applying it to the Milky Way using current sky surveys. Finally, Section \ref{disc} summarizes our findings and discusses the prospects of applying this framework to the Milky Way, along with the associated challenges and limitations.
\begin{figure*}[t!]
\centering
\includegraphics[width=\textwidth]{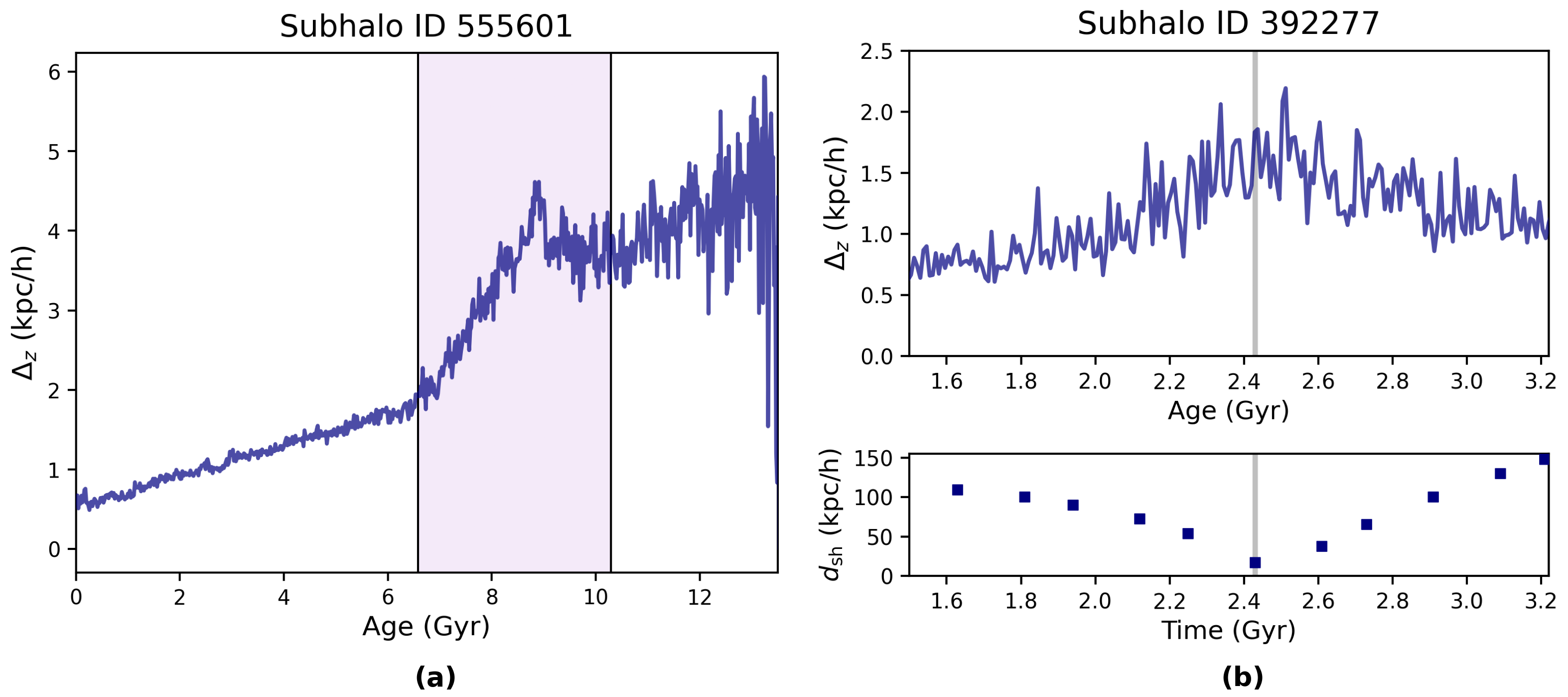}
\caption{(a) Age vs $\Delta_z$ trend of the present-day snapshot of the Milky Way analogue with ID 555601 in TNG50, featuring the signature of a major merger. The highlighted region marks a notable merger occurrence (see Figure \ref{fig:simulation_snap}), whose impact is evident in the vertical thickness of the stars formed during this period within the galaxy, as observed at present. (b)  Age vs $\Delta_z$ trend of the present-day snapshot of the model 392277 in TNG50 highlights a flyby's signature. The top right panel illustrates the influence of a flyby between 3.2 and 1.6 billion years ago on the thickness of stars formed during that time in the Milky Way analog 392277, as observed at present. Meanwhile, the bottom right panel depicts the distance between the satellite and the host galaxy ($d_{sh}$) over time (equivalent to the ages of the stars born during that time). The parameters in both panels show a significant correlation, with the maxima on the Age vs $\Delta_z$ plot in the top panel representing the time of closest approach of the flyby (marked in grey).}
\label{fig:fig2}
\end{figure*}
\section{Milky Way analogues from TNG50} \label{sec:tng50}

\begin{figure*}
\centering
\includegraphics[width=14cm]{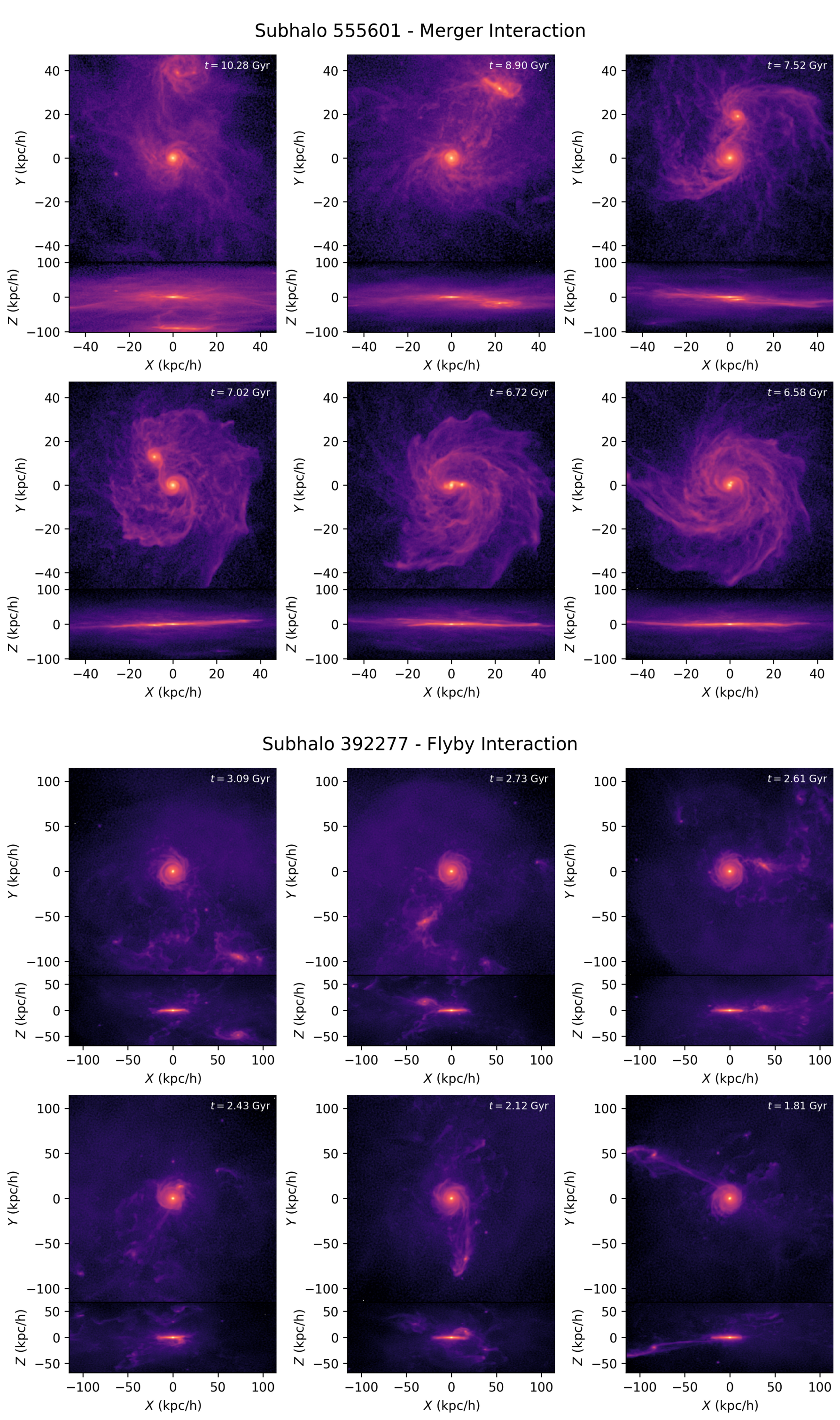}
\caption{Simulation snapshots of the interaction histories of Subhalo 555601 and Subhalo 392277. The snapshots are shown at lookback times comparable to the stellar age range associated with features in the age-thickness relation in Figure \ref{fig:fig2}.}

\label{fig:simulation_snap}
\end{figure*}
Cosmological simulations play a crucial role in understanding galaxy evolution and the impact of external disturbances on various galactic parameters. Among these, the IllustrisTNG suite \citep{Pillepich_2017, Springel_2017, Nelson_2017, Marinacci, Naiman, nels}, and especially its latest iteration TNG50 \citep{nels1}, offers high spatial and force resolution within a 50 Mpc box sampled by 2160$^3$ gas cells with a baryon mass resolution of $8.5 \times 10^4 M_{\odot}$. This study uses the publicly available MW/Andromeda analogs in TNG50 \citep{pillepich2023}. Using these simulations, we investigate how merger events are imprinted in the vertical structure of stellar disks. Throughout this work, simulation lengths are reported in physical kpc/h, adopting $h$=0.6774. We perform a coordinate transformation to rotate each galaxy such that the gaseous disk lies in the $x-y$ plane (face-on), with the $z$-axis perpendicular to the disk, using the gas within the inner $\sim 10$ kpc/h to define the midplane.

\subsection{Age-thickness relation}
As a representative example, we examine Subhalo 555601, a Milky Way analogue from TNG50, with a current total mass of 7.93×10$^{11}$M$_{\odot}$. To investigate the evolution of its disk thickness, we computed the relationship between stellar age and the dispersion in vertical positions by binning stellar ages into 700 bins (with a typical bin width of $\sim$ 0.02 Gyr, larger than the intrinsic age resolution of this subhalo snapshot). The vertical thickness of each stellar population is then quantified using the standard deviation of the vertical stellar positions in every age bin, given by: 
\begin{equation}
    \Delta_z = \sqrt{\frac{1}{n-1} \sum_{i} (z_i - \bar{z})^2},
\end{equation}
where n is the number of stars in each bin, $z_i$ is the vertical position of $i^{\rm th}$ star, and $\bar{z}$ is the mean of the vertical positions of all the stars in a given bin. Thus $\Delta_z$ measures the dispersion of stellar vertical positions about the mean plane of that bin. This quantity serves as a proxy for the vertical thickness of the stellar disk. In this analysis, we exclude stars with  $|z|>12$ kpc/h. This cut removes the high-$z$ stars that may be part of non-disk components. Although it changes the absolute normalization of the age–thickness relation, especially in the older age bins, repeating the analysis without this cut does not change the main conclusions presented here.  We verified that our results are insensitive to the precise choice of binning. Since $\bar{z}$ is computed independently for each age bin, large-scale vertical offsets such as warps do not contribute to the measured thickness. For Subhalo ID 555601, the Age--$\Delta_z$ relation is computed for stars located at intermediate galactocentric radii, whereas for Subhalo ID 392277, the relation is measured in the outer disk. These radial ranges were chosen because the signatures of the interactions were most prominent in these regions for the respective systems. The radial regions (inner, intermediate, and outer) are defined relative to the radial scale length of each galaxy model, as described below. A more detailed description of this classification, along with a comparison across different radial ranges within each galaxy model, is presented in section \ref{sec:rad}. 

Panel (a) of Figure \ref{fig:fig2} illustrates the relationship between stellar age and vertical thickness for Subhalo ID 555601. Older stellar populations generally display larger vertical thickness, consistent with a combination of birth thickness and subsequent dynamical heating. However, the distribution deviates significantly from this overall trend within the region highlighted in magenta. This deviation begins approximately 10.3 billion years ago, peaks at 8.9 billion years, and then declines sharply until 6.6 billion years ago, marking a period of increased vertical thickness. This period coincides with a merger event with a mass ratio of approximately 5:1. The thickening of the disk during this epoch arises from dynamical heating of the stellar disk and the disturbed state of the gaseous disk, triggered by gravitational perturbations from the merging satellite. The shaded region in Figure \ref{fig:fig2}a and the interaction snapshots of Subhalo ID 555601 shown in Figure \ref{fig:simulation_snap} highlight this phase of the interaction. We verified that the step-like feature remains present when the analysis is restricted to stars identified as in-situ in TNG50, although ex-situ stars contribute to the amplitude of the signal (see Appendix, Figure \ref{fig:A1}).

During the merger, gravitational perturbations increase the vertical motions of stars already present in the disk, leading to dynamical heating of pre-existing stellar populations \citep{toth,quinn}. As a result, stars older than the merger epoch exhibit enhanced vertical thickness. At the same time, the gaseous disk, being more extended and less gravitationally bound, is particularly susceptible to disruption, becomes dynamically disturbed, and more turbulent. Stars formed during this phase are born dynamically hot, producing a trend in the Age--$\Delta_z$ relation that starts at $\sim$ 10.3 and peaks at $\sim$ 8.9 Gyr as the satellite approaches the host and the gravitational perturbation becomes stronger. After the interaction, the gas dissipates energy and gradually settles back toward the disk plane. Because gas is collisional and dissipative, it can radiate away energy efficiently and settle into a dynamically cold configuration, from which new stars are born. The subsequent decline in $\Delta_z$ from 8.9 to 6.6 Gyr reflects the formation of progressively thinner stellar populations as the gaseous disk settles following the merger. By $\sim$ 6.6 Gyr, the gas has largely settled into a new disk plane as angular momentum is redistributed within the system, allowing star formation to proceed in a thin, dynamically cold disk \citep{donghia2006}. From this point onward, the galaxy evolves secularly, and the Age--$\Delta_z$ relation follows a smoother trend in the absence of further major mergers or strong perturbations. This leads to a clear separation between geometrically thick populations formed or affected during the merger and thinner populations formed afterward.

To investigate whether the present-day age--$\Delta_z$ feature arises solely from merger-driven heating of pre-existing stars or whether the progenitor disk already had an age-dependent vertical structure before the interaction, we examine the evolution of both the age--$\Delta_z$ and age--$\sigma_z$ relations before and after the merger (see Appendix \ref{sec:thickness_ev}). The age--$\sigma_z$ relation traces the vertical kinematic state of each stellar population, while the age--$\Delta_z$ relation reflects how those kinematics are mapped into a physical scale height by the evolving gravitational potential of the disk. The pre-interaction snapshots show a monotonic trend, with older populations already thicker and kinematically hotter than younger populations. In contrast, the post-interaction snapshots show enhanced thickness and a step-like feature associated with the merger epoch. Thus, the localized step remains an interaction signature, while the absolute amplitude of $\Delta_z$ reflects the combined effects of birth thickness, merger-driven and secular heating, radial redistribution, and the evolving vertical restoring force that maps vertical motions into a physical scale height.


In contrast, a flyby interaction produces a much weaker dynamical response in the stellar disk. As illustrated in the upper panel of Figure \ref{fig:fig2}b for Subhalo 392277, the Age--$\Delta_z$ relation in the present-day snapshot exhibits a modest peak for stellar populations between $\sim$ 1.6 and 3.2 Gyr. This peak indicates an increase in vertical dispersion during this period, corresponding to a flyby event starting approximately 3.2 Gyr ago. The lower panel shows the distance between the satellite and the host galaxy ($d_{\rm sh}$) as a function of time. The snapshots highlighting the flyby for this galaxy are shown in the bottom panel of Figure \ref{fig:simulation_snap}. The minimum in $d_{\rm sh}$ (around 17 kpc/h) at $\sim$2.4 Gyrs marks the time of closest approach of the satellite. The peak in $\Delta_z$ coincides with the minimum in $d_{\rm sh}$, indicating that variations in $\Delta_z$ can serve as a robust indicator of the timing of past gravitational encounters. In this case, the increase in $\Delta_z$ is primarily seen in stellar populations formed during the encounter, suggesting that the flyby primarily perturbs the gaseous disk, with only limited dynamical heating of the pre-existing stellar disk due to the shorter duration and weaker tidal forces of the interaction.


\subsection{Radial Variation of Age--$\Delta_z$ Signatures for Mergers and Flybys} \label{sec:rad}

\begin{figure*}[t!]
\centering
\includegraphics[width=\textwidth]{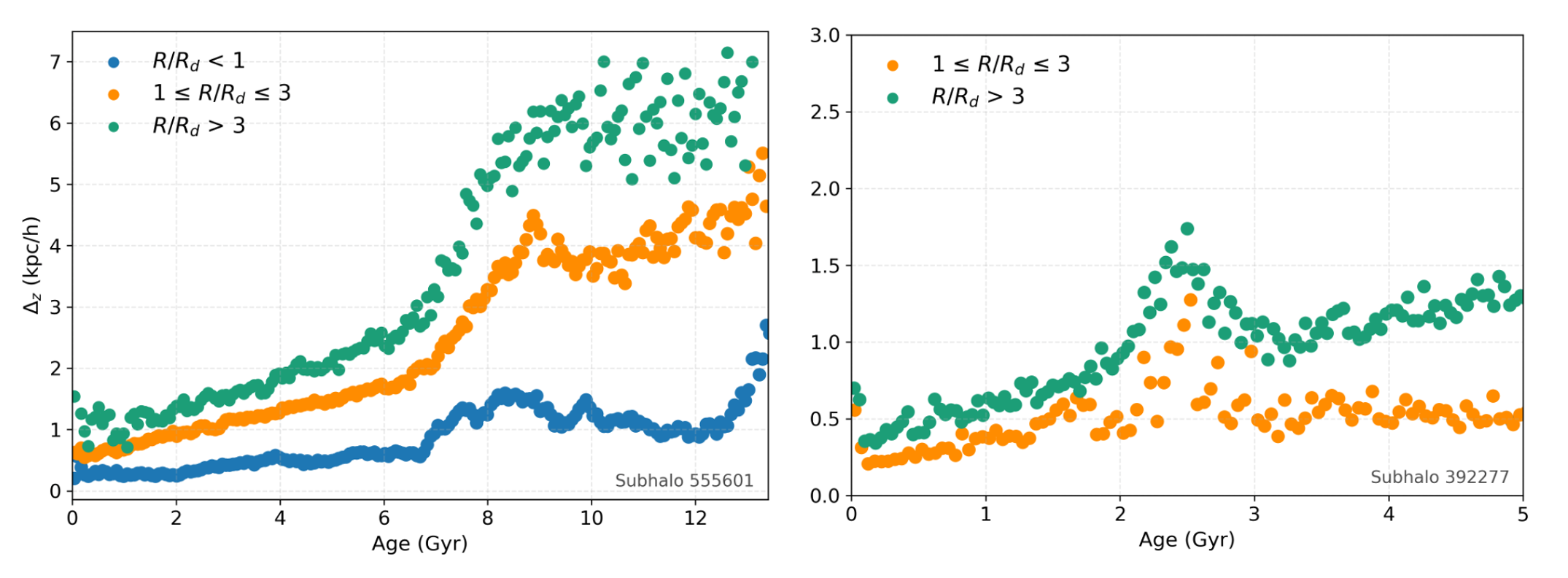}
\caption{Age--$\Delta_z$ relation in different radial regions (inner, intermediate, and outer) of the disk for the two Milky Way analogues. Panel (a) shows Subhalo ID 555601 with the major merger signature, whereas panel (b) shows Subhalo ID 392277 with the flyby signature. See section \ref{sec:rad} for a detailed description.}
\label{fig:fig3}
\end{figure*}
\begin{figure}
	\centering
    \includegraphics[width=8.5cm]{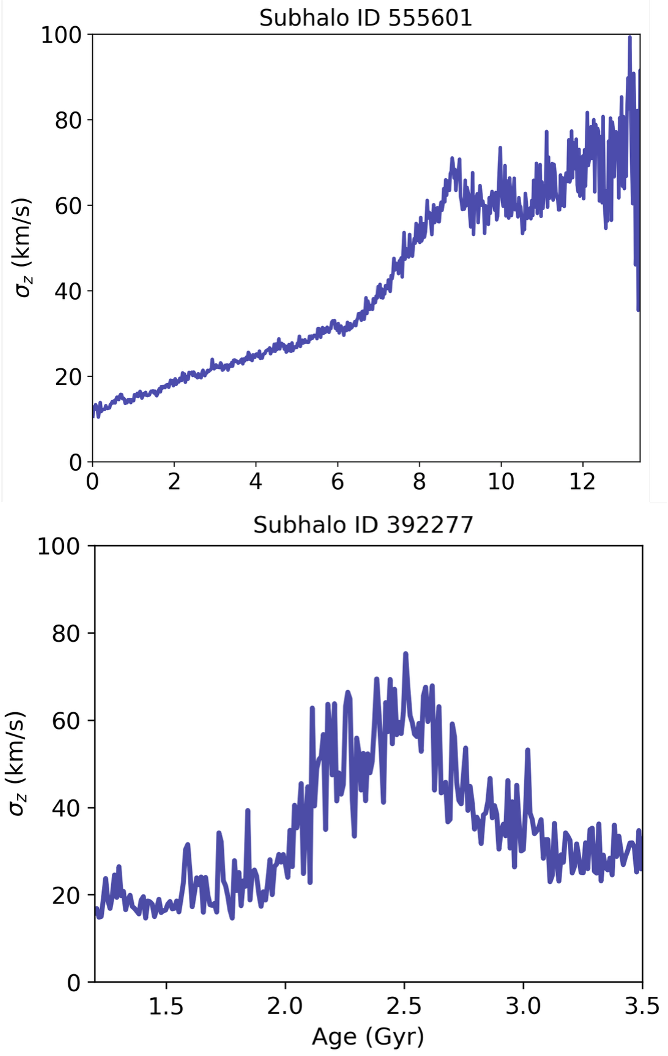}
    \caption{Age versus vertical velocity dispersion ($\sigma_z$) for stars in Subhalo 555601 (top panel) and Subhalo 392277 (bottom panel). The stars are selected from the same radial regions as in Figure~\ref{fig:fig2}, where the Age--$\Delta_z$ relation is shown (see Section \ref{sec:rad}).}
    \label{fig:fig7}
\end{figure}
To study the radial variation of the Age--$\Delta_z$ relation in each galaxy model, we divide the disk into inner, intermediate, and outer regions. This classification is defined relative to the scale length of the disk, $R_d$. We estimate $R_d$ by fitting an exponential profile to the stellar surface density distribution in the present-day snapshots, $\Sigma(R) \propto e^{-R/R_d}$. We then define the inner region as $R < R_d$, the intermediate region as $R_d \leq R \leq 3R_d$, and the outer region as $R > 3R_d$. For subhalo 555601, $R_d=8.2$ kpc/h, and for subhalo 392277, $R_d=2.5$ kpc/h. Figure \ref{fig:fig3}a and \ref{fig:fig3}b illustrate the radial dependence of the age--$\Delta_z$ relation for Subhalo ID 555601 and 392277, respectively. 

For Subhalo 555601, which experienced a major merger, a clear step-like feature in the Age--$\Delta_z$ relation is present in all three radial regions. However, the morphology of this feature varies with radius. The amplitude of vertical thickening increases with radius, consistent with the weaker vertical restoring force in the outer disk. In the inner disk, which is dominated by the bulge stars, the transition appears less pronounced and shows multiple small fluctuations, likely due to stronger dynamical mixing and a deeper gravitational potential in the central regions. In contrast, the outer disk shows a larger overall vertical response, but the lower stellar density in this region, together with the wide range of vertical excursions due to a weaker restoring force, makes the merger-onset feature less sharply defined. The intermediate region displays the clearest signature, where the peak marking the merger onset remains well-defined, allowing the merger epoch to be identified most clearly. To ensure that the large $\Delta_z$ values in Subhalo 555601 do not simply reflect a loss of disk-like structure or contamination by non-disk stellar components, we examine the morphology and kinematics of the same age-selected populations in Appendix \ref{app:555601_morphology}. These tests show that the merger-epoch and pre-merger epoch stellar populations remain flattened and rotating, consistent with a geometrically thicker disk component.

For Subhalo ID 392277, which experiences a flyby interaction, the age--$\Delta_z$ relation still shows an overall increase in amplitude with radius. However, in this case, the signature is strongest in the outer disk, although it remains visible at intermediate radii. This is expected because the flyby perturbs the gas in the outer disk more strongly, where the vertical restoring force is weaker, and stars formed from this gas inherit the resulting vertical structure. The inner disk of this galaxy contains no stars in the age range 0--5 Gyr, so this region is not shown in the figure.

\begin{figure*}[t!]
	\centering
    \includegraphics[width=18cm]{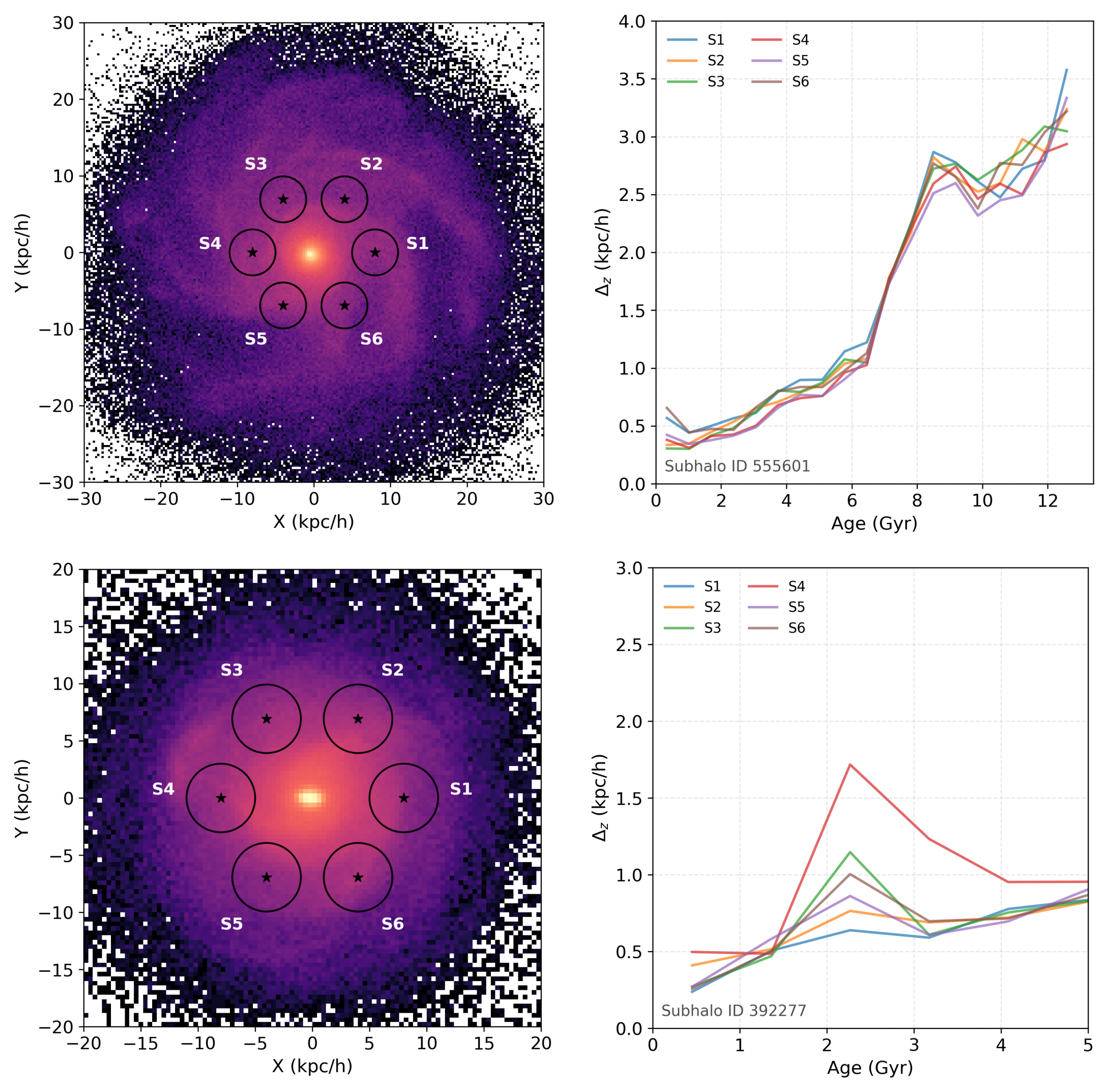}
    \caption{Age--$\Delta_z$ relation measured from the mock observer regions placed at different azimuthal locations around the disk in two Milky Way-like galaxies from TNG50. The top row corresponds to Subhalo ID 555601, which experienced a major merger, while the bottom row corresponds to Subhalo ID 392277, which underwent a flyby interaction. The left panels show the face-on stellar surface density maps, where six mock observers (S1--S6) are placed at a fixed galactocentric radius of $R=8$ kpc/h across varying azimuths in the disk. Around each location, stars within a cylindrical region of radius 3 kpc/h are selected to define a mock local observer volume. The right panels of both rows show the corresponding age--$\Delta_z$ curves measured in each of these regions.}
    \label{fig:fig4}
\end{figure*}
\subsection{Comparison with age--$\sigma_z$ Relation}
Figure \ref{fig:fig7} shows the Age versus vertical velocity dispersion ($\sigma_z$) relation for the two model galaxies, measured in the same regions as in Figure \ref{fig:fig2}. A comparison between Figures \ref{fig:fig2} and \ref{fig:fig7} shows that the age--$\Delta_z$ relation traces the same overall behaviour as the age--$\sigma_z$ relation. This similarity is expected because both quantities trace the vertical heating of the disk, although they rely on fundamentally different observables. For a detailed comparison of the age-velocity dispersion relation in simulations and observations, see \cite{mccluskey2025}.

The age--$\sigma_z$ relation has long been used, often combined with metallicities, to identify merger events in the Milky Way \citep[e.g.,][]{gustaf1946,nordstrom2004,payel,sun2025}. However, measuring $\sigma_z$ requires full 3D velocities, which are observationally more difficult to obtain. Consequently, analyses based on kinematics are limited to stars with reliable velocity measurements. In contrast, the age--$\Delta_z$ relation relies only on stellar positions and ages, and therefore does not require line-of-sight velocities nor proper motions. With current and upcoming surveys such as Gaia \citep{gaia2023}, Roman \citep{roman}, and LSST \citep{lsst2019} mapping stars over increasingly large volumes of the Milky Way, this approach allows the use of significantly large data from surveys where full kinematic information is incomplete or unavailable.

\section{Robustness of the age--$\Delta_z$ Signature to Observational Constraints}
\label{sec:robustness}
In current Galactic surveys, the best phase-space data and stellar ages are available near the Solar radius, approximately 8 kpc in the Milky Way \citep{reid2019}. It is therefore important to ask whether the age--$\Delta_z$ signature depends on where in the Galaxy we observe the stars. To test this, we place six mock observers at different azimuths around each disk at a fixed galactocentric radius of R=8 kpc/h (regions S1–S6). For each location, stars within a cylindrical region of radius 3 kpc/h  are selected, and the age–$\Delta_z$ relation is measured. Within each galaxy, the same age binning is used for all six locations, with 20 bins for Subhalo 555601 and 15 bins for Subhalo 392277. Age bins containing fewer than 100 stars are excluded from the analysis. The results are summarized in Figure \ref{fig:fig4}.

It is clear from the top row of Figure \ref{fig:fig4} that, despite the limited spatial sampling, Subhalo ID 555601 retains the major merger signature in the age--$\Delta_z$ relation across all six mock observer regions. All six regions trace a similar overall trend in the age--$\Delta_z$ relation, indicating that the ability to detect this signature is largely independent of the observer’s location for major mergers within the disk. This further supports the idea that the age--$\Delta_z$ relation is a reliable diagnostic for identifying past major merger interactions in the Milky Way.

The bottom row of  Figure \ref{fig:fig4}, which shows Subhalo ID 392277 with a flyby interaction, presents a different picture. In this case, whether or not the signature appears seems to depend on the location of the observer. Regions S3--S6 show a clear signature with larger amplitudes, with S4 displaying the largest amplitude, whereas regions S1 and S2  show only a weak response. This reflects the fact that a flyby does not perturb the entire disk uniformly, and regions closer to the perturber respond more strongly to the interaction than other parts of the disk. 
\begin{figure*}[t!]
	\centering
    \includegraphics[width=18cm]{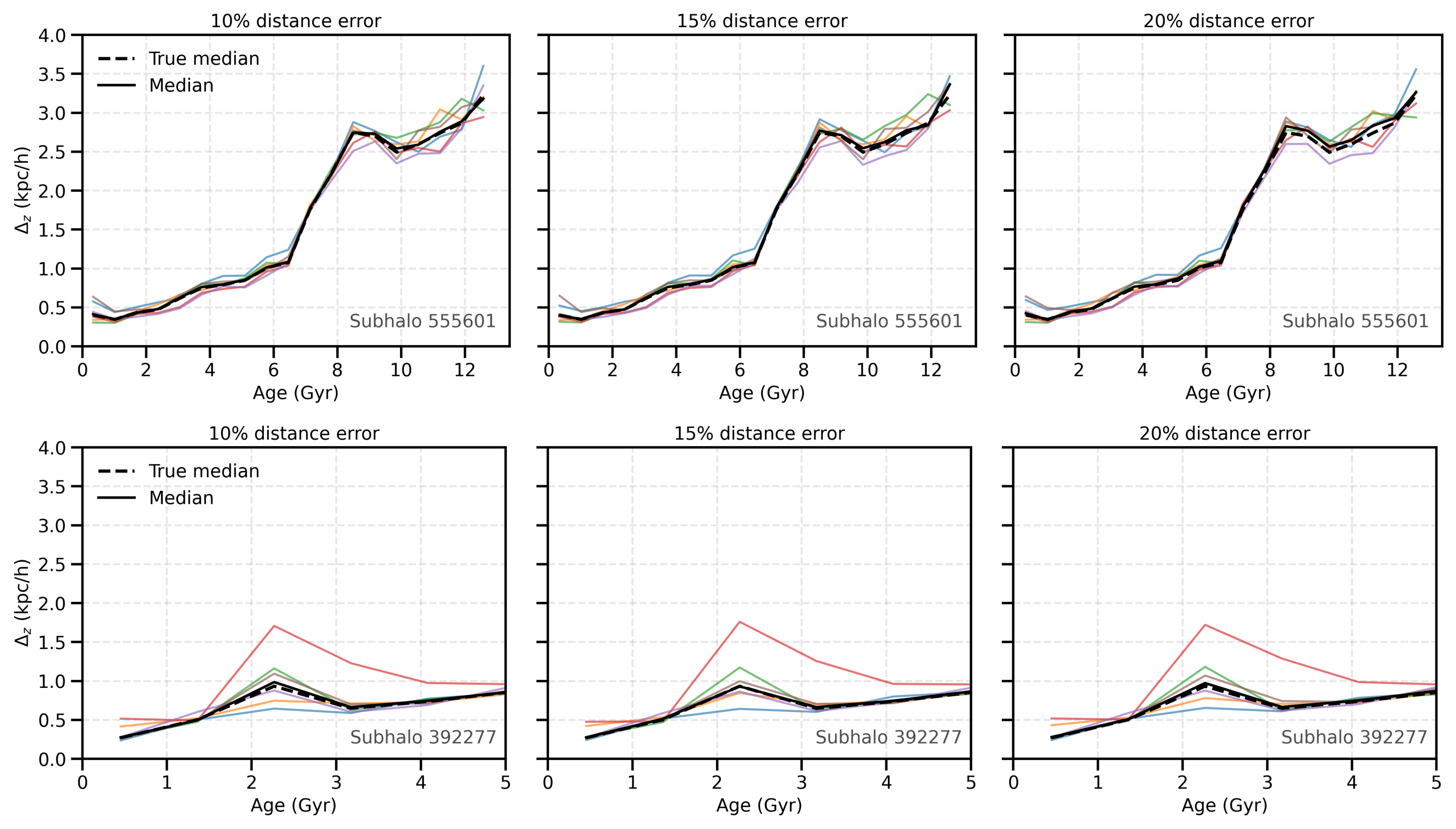}
     \caption{For each mock observer region S1--S6 from Figure \ref{fig:fig4}, we introduce fractional distance uncertainties of 10\%, 15\%, and 20\% and compare the median of the resulting curves (black line) with the true median from Figure \ref{fig:fig4} (dashed black line). The top panel shows the effect of distance uncertainties for Subhalo ID 555601, while the bottom panel shows the same for Subhalo ID 392277.}
    \label{fig:fig5}
\end{figure*}

While the azimuthal dependence of the signal might at first seem inconsistent with expectations from phase mixing \citep{antoja,ant1,hunt2022}, the two can in fact coexist. The age--$\Delta_z$ relation can persist as a long-lived imprint of the heating history, while any azimuthal dependence from a flyby should, in principle, fade as the system phase mixes. One natural explanation is that the interaction is recent enough that the outer disk has not yet fully mixed. In addition, a flyby is an intrinsically localized perturbation. Unlike a major merger, which redistributes energy and angular momentum across the entire disk, a flyby typically affects only a limited region of the galaxy. This naturally leads to azimuthal variations that are not global to begin with. Another key point is that the disk response to a flyby is not purely stochastic. The perturbation can excite coherent structures, such as spirals \citep[e.g.,][]{Gauthier2006,purcell2011,donghia2013}, warps \citep[e.g.,][]{ostriker1989,weinberg1995,poggio2021}, and bending modes \citep[e.g.,][]{purcell2011,gomez2013, widrow2014,laporte2018,hawthorn2021,thulasidharan2022} that are collective in nature and can persist for several orbital times, especially in the outer disk where dynamical times are long. In that sense, the azimuthal dependence may not simply reflect incomplete phase mixing, but rather a long-lived, spatially localized global response of the disk. Thus, while the age--$\Delta_z$ relation shows a weaker response to flybys farther from the region of closest interaction, it can still be useful when complemented with other diagnostics used to infer a galaxy’s merger history. 

That being said, spatial limitations are not the only challenge in real observations. We are also constrained by measurement uncertainties, particularly in stellar distances and ages. To account for this, we add uncertainties to the stellar ages and distances in each of the six mock observer regions and study how they affect the age--$\Delta_z$ relation. In addition, extinction leads to poor sampling near the midplane along the line of sight. We qualitatively study its impact on the inferred age--$\Delta_z$ relation by masking sources close to the midplane. 
\begin{figure*}
	\centering
    \includegraphics[width=18cm]{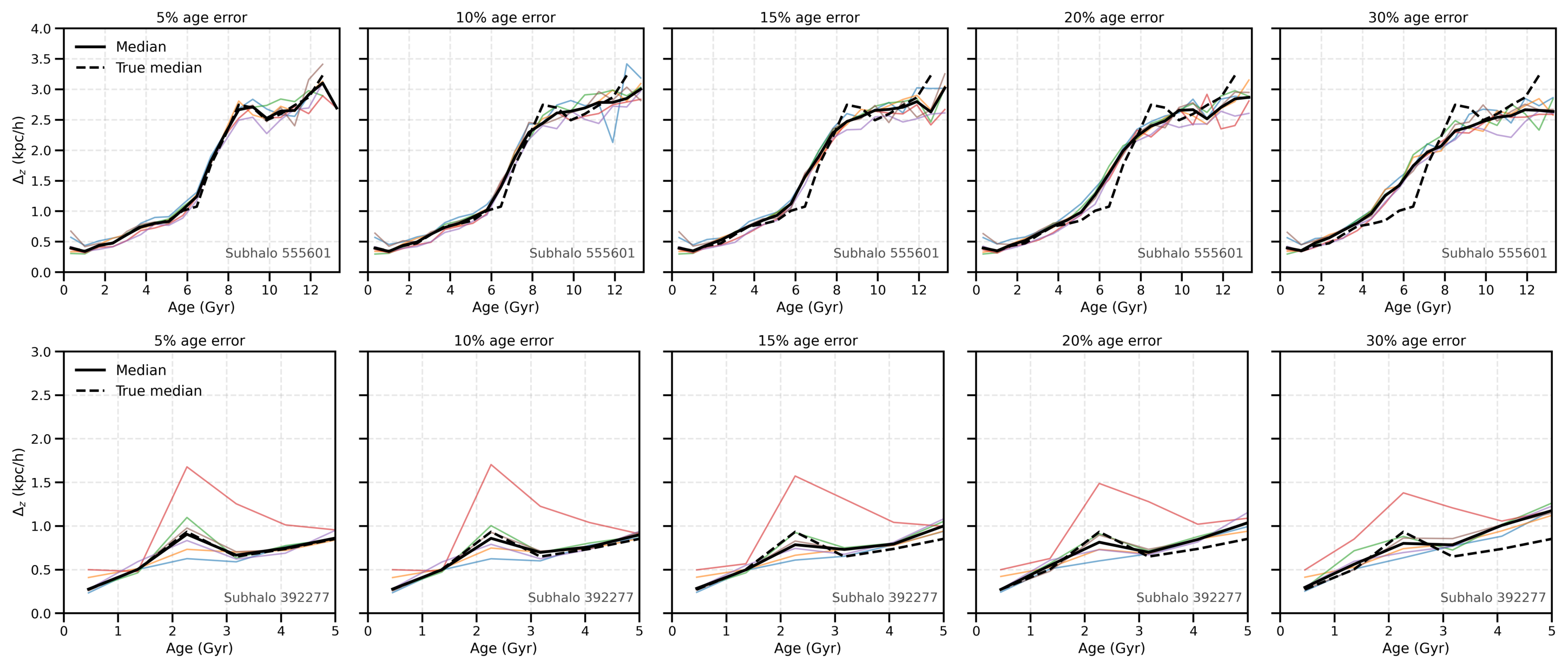}
     \caption{For each mock observer region S1--S6 from Figure \ref{fig:fig4}, we introduce fractional uncertainties in stellar ages of 5\%, 10\%, 15\%, 20\%, and 30\% and compare the median of the resulting curves (black line) with the true median from Figure \ref{fig:fig4} (dashed black line). The top panel shows the effect of age uncertainties for Subhalo ID 555601, while the bottom panel shows the same for Subhalo ID 392277.}
    \label{fig:fig6}
\end{figure*}
\subsection{Effect of Distance Uncertainties}
\label{sec:dist_error}
To model distance uncertainties, we assume a fractional Gaussian error in the stellar distance, $d'=d(1+\delta)$, where $\delta$ is the fractional error in distance and $\delta \sim \mathcal{N}(0,\sigma_d^2)$. The vertical coordinate $z$ is related to the distance through
$z=d\sin b$, where $b$ is the Galactic latitude. Propagating this uncertainty gives $z'=d'\sin b= z(1+\delta)$.  We then recompute the age--$\Delta_z$ relation for regions S1-S6 assuming fractional Gaussian uncertainties of 10\%, 15\%, and 20\% in the stellar distances for both model galaxies. The results are summarized in Figure \ref{fig:fig5}.

We compare the median of the S1--S6 curves (black line) with the corresponding median from Figure~\ref{fig:fig4} (dashed black line), which represents the true median, to assess the effect of distance uncertainties on the age--$\Delta_z$ relation. The plots show that distance uncertainties at the level of 10\%, 15\%, or 20\% do not significantly affect the observed age--$\Delta_z$ relation. However, note that since the uncertainties we model are random, their dominant effect is to broaden the age--$\Delta_z$ relation. Any systematic bias should therefore be interpreted with caution \citep{sch2019}. Also note that the distance uncertainties are Gaussian in parallax and thus asymmetric in distance \citep{luri2018,Bailer_Jones_2021}, especially for fractional uncertainties larger than $20\%$ \citep{BaileJones2015}. Our implementation assumes symmetric errors, making it a simplified, somewhat optimistic case. However, in this work we only consider ``good quality'' distance measurements (fractional uncertainties $\leq 20\%$) for which the asymmetry remains reasonably small, and their treatment therefore provides a reasonable approximation.

\begin{figure*}[t!]
	\centering
    \includegraphics[width=\textwidth]{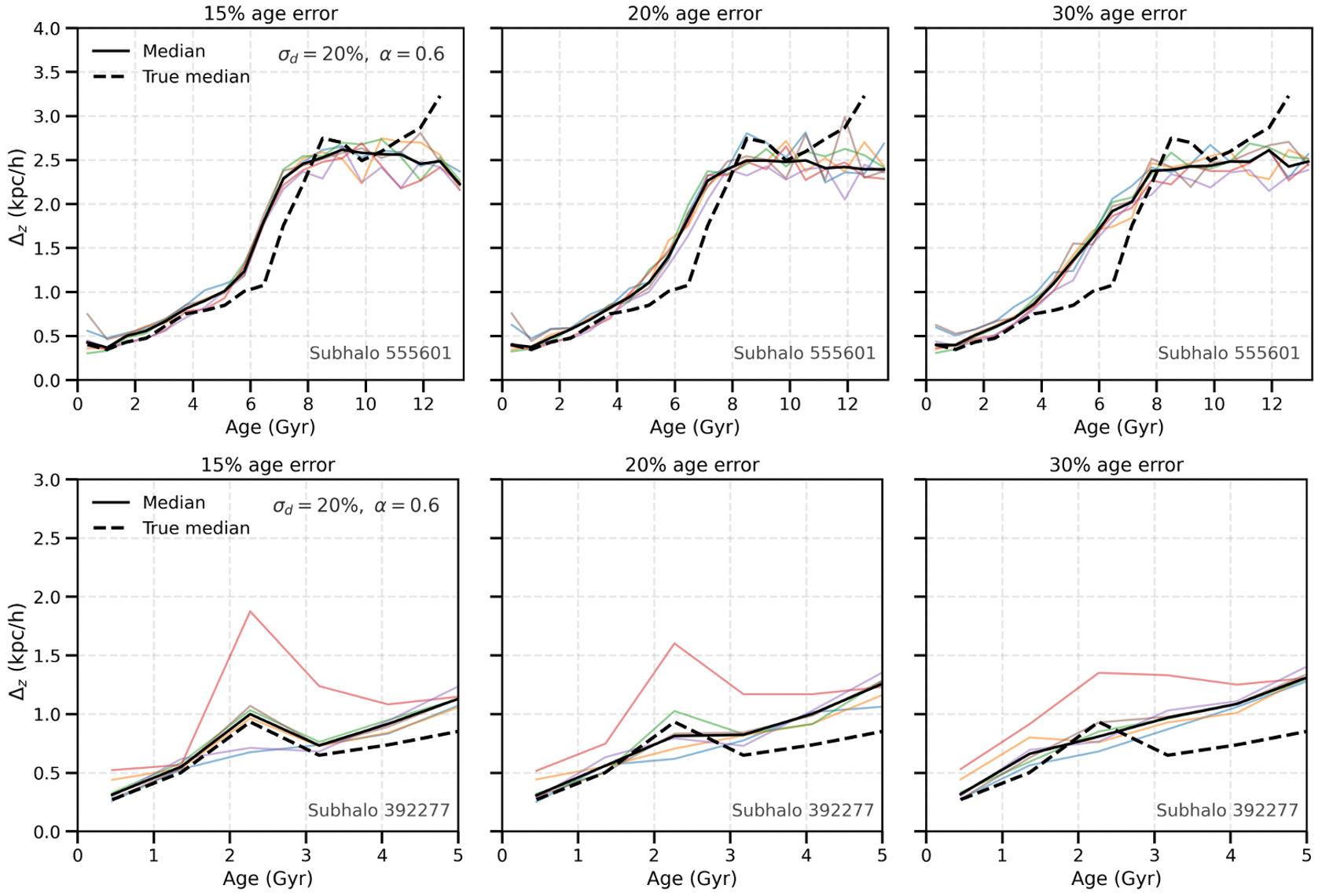}
   \caption{For each mock observer region S1--S6 from Figure \ref{fig:fig4}, we apply the same distance errors to both the vertical positions and the stellar ages, so that the two are correlated. We assume a fixed fractional distance uncertainty $\delta=20\%$ and a strength parameter $\alpha=0.6$, and adjust the remaining independent Gaussian scatter in log-age to achieve total age uncertainties of 15\%, 20\%, and 30\% (see section \ref{sec:corr} for more details ). The solid black line shows the median of the resulting curves, while the dashed black line shows the true median from Figure \ref{fig:fig4}. The top panel corresponds to Subhalo ID 555601, and the bottom panel shows the same for Subhalo ID 392277.}
    \label{fig:corr}
\end{figure*}

\subsection{Effect of Uncertainties in Stellar Ages}
\label{sec:age_error}
Stellar ages are notoriously difficult to measure. Modern techniques such as isochrone fitting \citep{fred2004,jorgensen2005,Feuillet2018,xian,nataf2024}, asteroseismology \citep{chaplin2013}, and, more recently, machine-learning-based approaches \citep{lam2,que,Almannaei2026} have made it possible to estimate stellar ages for millions of stars; however, the resulting age uncertainties remain substantial. Some catalogs report age uncertainties as low as $\sim$ 7\% \citep[e.g.,][]{nataf2024}, whereas others report uncertainties as high as $\sim$ 30\% \citep[e.g.,][]{lam2}.

To study the effect of uncertainties in stellar ages on the age--$\Delta_z$ relation, we introduce age uncertainties in regions S1--S6 for both model galaxies. We adopt log-normal errors to account for the fact that age uncertainties in real observational datasets are often asymmetric. We model the observed age as $\tau'=\tau \exp(\eta)$, where $\eta\sim\mathcal{N}(0,\sigma_{\ln \tau}^2)$. The width of the distribution is chosen so that the 84$^{th}$ percentile corresponds to the adopted fractional age uncertainty, i.e., $\sigma_{\ln \tau}=\ln(1+\epsilon)$,
where $\epsilon$ is the fractional error. We then recompute the age--$\Delta_z$ relation for each of the six mock observer regions for both Subhalo IDs 555601 and 392277, assuming fractional age uncertainties of 5\%, 10\%, 15\%, 20\%, and 30\%, reflecting the range of uncertainties reported in the literature. The results are summarized in Figure \ref{fig:fig6}.

For Subhalo 555601 (top row), we compare the median of the S1--S6 curves (black line) with the true median from Figure~\ref{fig:fig4} (dashed black line). This comparison shows that age uncertainties significantly affect the observed age--$\Delta_z$ relation. While 5\% uncertainties have little impact, once the age uncertainties reach 10\%, the well-defined peak marking the onset of merger-driven heating starts to broaden. The effect of age uncertainties is more pronounced for older stars. Even at the 10\% level, the median curve for stars older than $\sim6$ Gyr starts to deviate from the true median, and this deviation increases with larger age uncertainties. At 30\% uncertainties, deviations are noticeable even for stars younger than $\sim6$ Gyr.
Although the overall signature of the merger event remains visible across all panels, the broadening of the peak makes it difficult to determine the merger epoch precisely. This implies that age uncertainties must be $\leq$ 10--15\% to reliably recover the timing of merger events from the age--$\Delta_z$  relation.

\begin{figure}
    \centering 
    \includegraphics[width=8.5 cm]{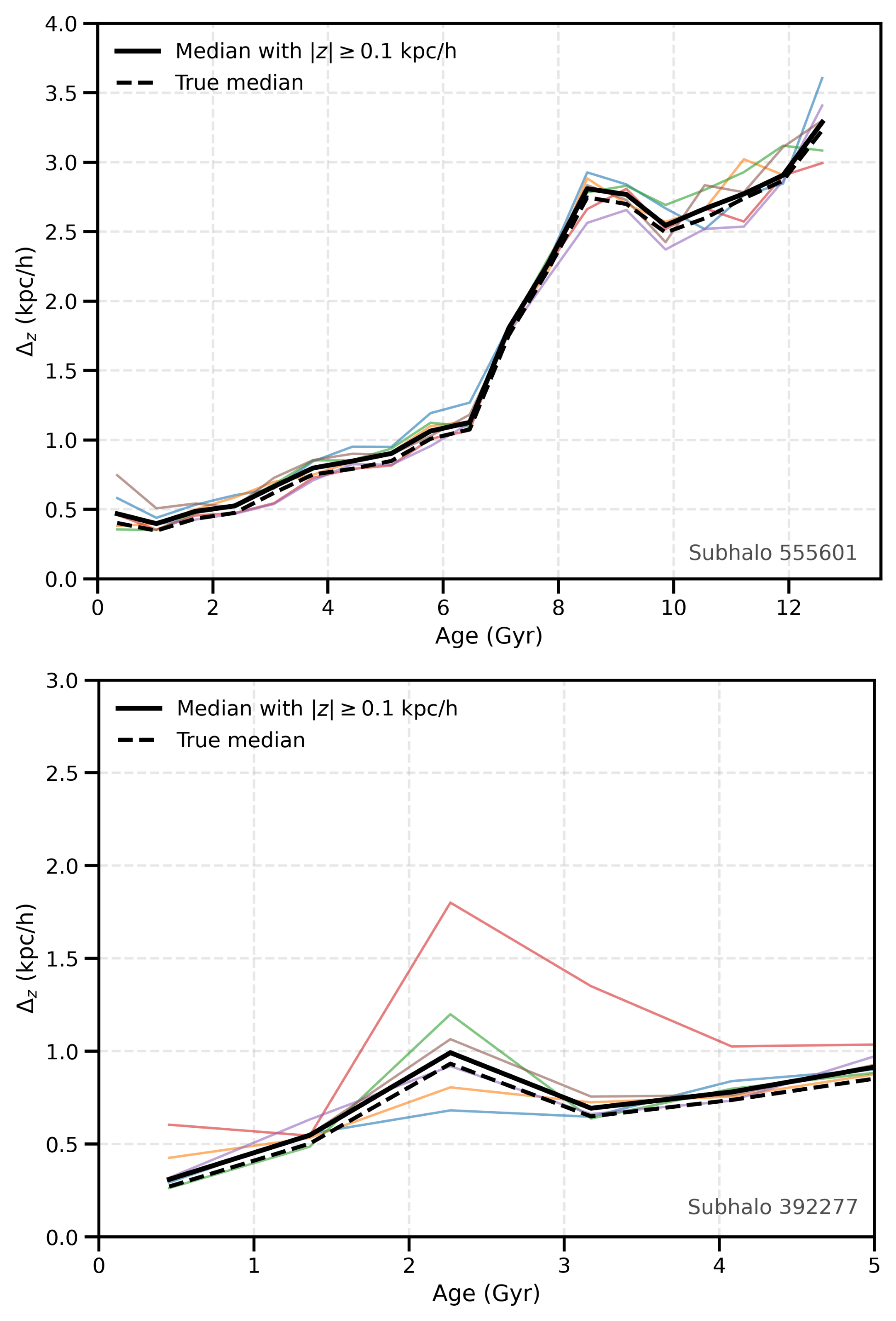}
    \caption{For each mock observer region S1--S6 from Figure \ref{fig:fig4}, we mask the region $|z|\leq 0.1$ kpc/h as a simple toy model to mimic the loss of stars close to the midplane due to extinction and compare the median of the resulting curves (black line) with the true median from Figure \ref{fig:fig4} (dashed black line).}
    \label{fig:fig8}
\end{figure}

Subhalo 392277 (bottom row) shows a similar trend, although the focus here is on younger stars because the flyby occurred more recently. In this case, the observed signal also depends strongly on the observer’s location within the galaxy, since the amplitude of the response varies across the different regions. In the figure, the median curve generally follows most of the regional curves, with the exception of region S4 (red curve). As the age uncertainties increase, the median begins to deviate from the true median even for very young stars. The peak associated with the flyby also broadens, although the flyby signature can still be distinguished in the median curve. Looking at the individual curves for regions S1--S6 reveals more information. In region S4 (red curve), which shows the strongest response before introducing age uncertainties, the signal remains visible even with 30\% age uncertainties. In region S1 (blue curve), which shows the weakest response without age uncertainties, the signal is completely lost at the 30\% uncertainty level. This suggests that for flyby events, the detectability of the signal depends strongly on the observer’s location in the disk, especially in the presence of age uncertainties, unlike the merger case, where the signal remains coherent across regions. Note that when introducing age uncertainties, we applied them across the full age range (0--13.6 Gyr) for this galaxy model as well, even though the plots shown here focus on 0--5 Gyr to highlight the flyby signal. We also repeated the analysis assuming Gaussian age errors for both galaxies and found the same qualitative behaviour.
\subsection{Effect of Correlated Distance--Age Uncertainties}
\label{sec:corr}

In the previous two sections, we studied the effects of fractional distance uncertainties and age uncertainties, treating them as independent quantities. In reality, both are present simultaneously, and stellar age and distance uncertainties are correlated. The absolute magnitude depends on the distance of a source through the distance modulus relation. If the distance is overestimated, the inferred absolute magnitude will be too bright, resulting in an underestimated stellar age. On the other hand, if the distance is underestimated, the inferred age will be overestimated.

In the Milky Way stellar age catalogues, the reported age uncertainties include contributions from several factors, only one of which is the distance uncertainty. We therefore use the term $-\alpha \delta$ to represent the contribution of distance errors,  where $\alpha$ quantifies the strength of the effect of distance uncertainties on the inferred stellar ages and $\delta$ is the fractional distance error, as defined in section \ref{sec:dist_error}. The negative sign ensures a negative correlation between distance and inferred age. All other sources of uncertainty are captured by the term $\eta$, which we model as a Gaussian scatter in log-age, as described in section \ref{sec:age_error}. We combine these contributions by modeling the observed age as  $\tau' = \tau \exp(\eta - \alpha \delta)$.
We then study the effect of both distance uncertainties and correlated distance--age uncertainties for total age uncertainties of 15\%, 20\%, and 30\%, where part of the uncertainty arises from distance errors and the rest from independent Gaussian scatter in log-age. We simultaneously propagate the distance uncertainty to the vertical positions following the same procedure as described in Section~\ref{sec:dist_error}. 

We adopt $\sigma_d = 20\%$, corresponding to our upper limit for ``good'' distances, and $\alpha = 0.6$, which ensures that distance errors contribute a reasonable fraction of the total age uncertainty. For example, for a total age uncertainty of 15\%, approximately half of the contribution arises from distance uncertainties. However, for these choices, total age uncertainties of 5\% and 10\% are not achievable, as the contribution from distance errors alone is larger than these thresholds. 

Our results are summarized in Figure \ref{fig:corr}. Comparing the true median from Figure \ref{fig:fig4} (dashed black line) with the median of regions S1--S6 (solid black line) shows that, for the chosen parameters, the inferred age--$\Delta_z$ relation is systematically shifted towards younger ages, particularly for older stellar populations, in addition to the broadening of peaks seen in Figure \ref{fig:fig6}. This effect is significant even at the 15\% total age uncertainty level. This behavior is most evident in Subhalo 555601, where the step-like signature associated with a major merger is shifted to younger ages relative to the true median, leading to an underestimate of the merger time. In contrast, the flyby signature shows behavior similar to that in Figure \ref{fig:fig6}. This is because the flyby is a more recent event, and its signature lies in younger stars, where the shift in age is smaller, and the dominant effect is the broadening of peaks. 

We note that our choice of fractional distance uncertainty of 20\% is somewhat pessimistic. We experimented with other values of $\sigma_d$, and found that the shift in older stars becomes more and more pronounced as the uncertainty increases. Even at a 10\% fractional uncertainty in distances, the shift remains prominent. This analysis therefore underscores the importance of having accurate distance estimates for studying the merger history of the Milky Way.

\subsection{Effect of Extinction--Induced Incompleteness}

The data from Galactic surveys such as Gaia are affected by selection effects such as the one due to interstellar extinction \citep{massa1989,cardelli1989,mathis1990}. The presence of dust along the line of sight leads to poor sampling very close to the midplane. In the context of our analysis, if we were to examine the distribution of stellar ages versus vertical positions of the stars in real data, we would expect a lack of sources near the midplane. To mimic this effect, as a simplified proxy, we mask stars within $|z|\leq0.1$ kpc/h in each of the mock observer regions S1--S6. This provides a qualitative estimate of how incompleteness near the disk plane may bias the inferred age--$\Delta_z$ relation. Note that our goal is not to model extinction realistically, which depends on several factors such as line of sight, distance, latitude, etc., but rather to capture its primary consequence: the loss of stars near the disk plane. We then compare the true median with the median computed after applying this mask. The results are shown in Figure \ref{fig:fig8}. Our analysis indicates that midplane masking preserves merger signatures but systematically overestimates the disk thickness. Thus, although the overall trend remains robust, the absolute thickness values show a bias towards higher numbers.

\section{Discussion and Conclusion}
\label{disc}
Our analysis shows that galaxies can preserve signatures of merger events as variations in the vertical thickness of their stellar disks. Merger and flyby events are followed by an increase in the vertical dispersion of stellar positions $\Delta_z$, serving as a proxy for disk thickness.

The influence of satellites and mergers on the heating of a galaxy's disk has been previously proposed through various simulation studies \citep{toth,quinn,donghia2010, Martig_2014,Grand_2016,Khoperskov_2023,Liang2026}. In this work, using the TNG50 simulations, we show that the age--$\Delta_z$ relation provides an observationally accessible diagnostic for probing the merger history of Milky Way-like galaxies and can serve as a complementary approach to traditional methods that rely on stellar kinematics and chemical abundances. In particular, we demonstrate that major mergers and flybys leave distinct signatures in the age--$\Delta_z$ relation. Major mergers produce a step-like transition in the age--$\Delta_z$ relation across the disk, separating a geometrically thick, pre-merger population with a larger $\Delta_z$ from a geometrically thin, post-merger population with smaller $\Delta_z$. The post-merger evolution of this relation therefore provides an estimate of the timescale over which the disk resettles into a new equilibrium plane and begins forming a thinner disk. However, the larger $\Delta_z$ of the geometrically thicker disk component should not be interpreted as a measure of merger-driven heating alone, but as the combined outcome of birth thickness, merger-driven and secular heating, radial redistribution, and the evolving vertical restoring force of the disk. Flybys, on the other hand, generate more localized peaks in the relation, whose amplitudes vary across the disk depending on which regions are most strongly affected by the interaction. 

Recent JWST observations have suggested that galactic disks assemble upside-down, with thick disk components forming before thinner embedded disks \citep{LianLuo2024, Tsukui2025}. These studies provide a statistical view of disk thickness evolution by comparing disk galaxies at different redshifts. Our analysis is complementary, as it uses the present-day age--$\Delta_z$ relation of an individual galaxy as a fossil record of its vertical assembly history. A smooth monotonic age--thickness relation in the present-day disk of a galaxy alone cannot distinguish between stars that were born thick and stars that were heated later. However, the merger-associated step-like feature shown in Figure \ref{fig:fig2}a provides an additional diagnostic. If a similar step-like signature separating geometrically thick and thin disk populations is observed in the age--$\Delta_z$ relation of the Milky Way, especially with a corresponding signature in the age--$\sigma_z$ relation, it may provide evidence for upside-down disk growth in the Galactic disk.      

With the availability of extensive data on Milky Way stars over increasingly larger volumes and improved stellar age estimates, we can potentially constrain the merger history of the Milky Way by examining the relationship between stellar ages and the dispersion in vertical positions. This approach has an observational advantage because it relies only on stellar positions and distances, in addition to ages. In contrast, traditional diagnostics based on kinematics often require 3D velocity information, which is harder to obtain for large samples.

Having said that, while mergers play a significant role in shaping the age--$\Delta_z$ relation, other processes such as bar-driven resonances \citep{saha2010,Grand_2016}, spiral arms \citep{medina2015}, scattering by giant molecular clouds (GMCs) \citep{spitzer1953,lacey1984,haninnen2002}, and clustered star formation \citep{pavel2002,pavel2011} can also contribute to disk heating in the vertical direction. Some of these processes, such as scattering by GMCs, may not be fully resolved in cosmological simulations such as TNG50. Therefore, the age--$\Delta_z$ relation should be treated as a complementary diagnostic and used together with other methods when inferring the merger history of the Milky Way. While the Age–$\Delta_z$ relation is directly observable, its physical interpretation should ultimately be tested across different galaxy formation models.

Applying this framework to the Milky Way will require careful consideration of observational limitations, including selection effects, spatial incompleteness, and uncertainties in stellar ages and distances. Combining Gaia data \citep{gaia2023} with spectroscopic surveys such as APOGEE \citep{apogee} and future datasets from Roman \citep{roman2024,roman} and LSST \citep{lsst2019} will be crucial for maximizing the diagnostic power of this approach.
In an earlier, pre-publication version of this investigation \citep{thulasidharan2024}, we reported a possible detection of signatures of merger- or flyby-driven perturbations in the Milky Way using stellar ages from a LAMOST-based sample of stars \citep{lam2} along with starhorse ages of metal-poor stars from SDSS DR12 \citep{que2018,que}.  Since this work, several additional datasets reporting stellar ages have become available to further explore this diagnostic \citep[e.g.,][]{nataf2024,Almannaei2026}. In a companion paper (Thulasidharan et al. 2026 (in prep)), a comparison of the models presented here to these data will be addressed.  

The age--$\Delta_z$ relation provides a physically intuitive link between the vertical structure of stellar populations and the dynamical history of disk galaxies. Because the age--$\Delta_z$ relation is sensitive to both the heating of pre-existing stellar populations and the stellar birth conditions, it offers a useful diagnostic of past interactions. As upcoming sky surveys provide increasingly precise and extensive data, stellar age estimates and distance measurements will continue to improve. This framework can therefore serve as a complementary tool for identifying signatures of merger and flyby events, and more broadly for interpreting the dynamical evolution of disk galaxies across cosmic time.






%
\begin{acknowledgments}
\textbf{Acknowledgments:} 
E.D acknowledges funding support from these grants: HST Cycle 30, HST-AR-17053.004. RD and EP were supported in part by the Italian Space Agency (ASI) through contract ASI-INAF 2025-10-HH.0 to the National Institute for Astrophysics (INAF). We thank the anonymous referee for their valuable comments and suggestions. 
\end{acknowledgments}

\bibliography{sample631}

@ARTICLE{BaileJones2015,
       author = {{Bailer-Jones}, Coryn A.~L.},
        title = "{Estimating Distances from Parallaxes}",
      journal = {\pasp},
         year = 2015,
        month = oct,
       volume = {127},
       number = {956},
        pages = {994},
          doi = {10.1086/683116},
archivePrefix = {arXiv},
       eprint = {1507.02105},
 primaryClass = {astro-ph.IM},
       adsurl = {https://ui.adsabs.harvard.edu/abs/2015PASP..127..994B}
}

@ARTICLE{hunt2022,
       author = {{Hunt}, Jason A.~S. and {Price-Whelan}, Adrian M. and {Johnston}, Kathryn V. and {Darragh-Ford}, Elise},
        title = "{Multiple phase spirals suggest multiple origins in Gaia DR3}",
      journal = {\mnras},
         year = 2022,
        month = oct,
       volume = {516},
       number = {1},
        pages = {L7-L11},


}

@ARTICLE{white1,
   author = { White, S. D. M. and {Rees}, M.~J.},
   title = {{Core condensation in heavy halos: a two-stage theory for galaxy formation and clustering}},
   journal ={MNRAS},
   volume = {183},
   pages = {341-358},
   year = {1978}
}

@ARTICLE{white,
   author = { White, S. D. M. and Frenk, C. S.},
   title = {{ Galaxy Formation through Hierarchical Clustering }},
   journal = {Astrophysical Journal},
   volume = {379},
   pages = {52},
   year = {1991}
}

@ARTICLE{kaufmann,
   author = { Kauffmann, G. and White, S. D. M. and Guiderdoni, B. },
   title = {{ The formation and evolution of galaxies within merging dark matter haloes.}},
   journal = {MNRAS},
   volume = {264},
   pages = {201-218},
   year = {1993}
}

@ARTICLE{helmi1,
   author = { Helmi, A. and White, S. D. M. },
   title = {{ Building up the stellar halo of the Galaxy.}},
   journal = {MNRAS},
   volume = {307},
   pages = {495-517},
   year = {1999}
}

@ARTICLE{donghia2010,
       author = {{D'Onghia}, Elena and others},
        title = "{Quasi-resonant Theory of Tidal Interactions}",
      journal = {\apj},
         year = 2010,
        month = dec,
       volume = {725},
       number = {1},
        pages = {353-368},
          
}

@ARTICLE{donghia2006,
       author = {{D'Onghia}, Elena and {Burkert}, Andreas and others},
        title = "{How galaxies lose their angular momentum}",
      journal = {\mnras},
         year = 2006,
        month = nov,
       volume = {372},
       number = {4},
        pages = {1525-1530},

}

@ARTICLE{gaia,
   author = {{Gaia Collaboration.}},
   title = {{Gaia Data Release 2. Summary of the contents and survey properties.}},
   journal = {A\&A},
   volume = {616},
   pages = {A1},
   year = {2018}
}

@ARTICLE{apogee,
   author = {Majewski, S. R. and others},
   title = "{ The Apache Point Observatory Galactic Evolution Experiment (APOGEE)}",
   journal = {Astron. J.},
   volume = {154},
   pages = {94},
   year = {2017}
}

@ARTICLE{helmi,
   author = {Helmi, A. and others},
   title = {{The merger that led to the formation of the Milky Way’s inner stellar halo and thick disk}},
   journal = {Nature},
   volume = {563},
   pages = {85-88},
   year = {2018}
}

@ARTICLE{bel,
   author = {Belokurov, V. and others},
   title = {Co-formation of the disc and the stellar halo},
   journal = {MNRAS},
   volume = {478},
   pages = {611-619},
   year = {2018}
}

@article{Belokurov_2022,
   title={Energy wrinkles and phase-space folds of the last major merger},
   volume={518},
   journal={MNRAS},
   author={Belokurov, Vasily and others},
   year={2022},
   pages={6200–6215} }

@ARTICLE{gilmore,
   author = {Gilmore, G. and Wyse, R. F. G. and Norris, J. E.},
   title = {{Deciphering the Last Major Invasion of the Milky Way.}},
   journal = {ApJ},
   volume = {574},
   pages = {L39-L42},
   year = {2002}
}

@ARTICLE{navarro,
   author = {Navarro, J. F. and Helmi, A. and Freeman, K. C.},
   title = {{The Extragalactic Origin Of The Arcturus Group}},
   journal = {ApJ},
   volume = {601},
   pages = {L43-L46},
   year = {2004}
}

@ARTICLE{xian,
   author = {Xiang, M. and Rix, HW.},
   title = {{A time-resolved picture of our Milky Way’s early formation history.}},
   journal = {Nature},
   volume = {603},
   pages = {599-603},
   year = {2022}
}

@ARTICLE{mont,
   author = {Montalbán, J. and others },
   title = {{Chronologically dating the early assembly of the Milky Way}},
   journal = {Nature Astronomy},
   volume = {5},
   pages = {640-647},
   year = {2021}
}

@ARTICLE{ciuca,
   author = {Ciucă, I. and Kawata, D. and Ting, Y. and Grand, R. J. J. and Miglio, A. and Hayden, M. and Baba, J. and Fragkoudi, F. and Monty, S. and Buder, S. and Freeman, K.},
   title = {{Chasing the impact of the Gaia-Sausage-Enceladus merger on the formation of the Milky Way thick disc}},
   journal = {MNRAS Letters},
   volume = {528},
   pages = {L122},
   year = {2024}
}

@ARTICLE{bel1,
   author = {Belokurov, V. and others},
   title = {The biggest splash},
   journal = {MNRAS},
   volume = {494},
   pages = {3880-3898},
   year = {2020}
}

@article{Malhan_2022,
   title={The Global Dynamical Atlas of the Milky Way Mergers: Constraints from Gaia EDR3–based Orbits of Globular Clusters, Stellar Streams, and Satellite Galaxies},
   volume={926},
   journal={ApJ},
   author={Malhan, Khyati and others},
   year={2022},
   pages={107} }

@ARTICLE{antoja,
   author = {Antoja,T. and others},
   title = {A dynamically young and perturbed {Milky Way }disk},
   journal = {Nature.},
   volume = {561},
   pages = {360–362},
   year = {2018}
}

@ARTICLE{ant1,
   author = {Antoja, T. and Ramos, P. and García-Conde, B. and Bernet, M. and Laporte, C. F. P. and Katz, D.},
   title = {The phase spiral in{ Gaia DR3}},
   journal = {MNRAS},
   volume = {673},
   pages = {1501–1506},
   year = {2020}
}

@ARTICLE{thulasidharan2022,
   author = {Thulasidharan, L. and D’Onghia, E. and Poggio, E. and Drimmel, R. and Gallagher III, J. S. and Swiggum, C. and Benjamin, R. A. and Alves, J.},
   title = {Evidence of a vertical kinematic oscillation beyond the {Radcliffe wave}},
   journal = {A\&A},
   volume = {660},
   pages = {L12},
   year = {2022}
}

@ARTICLE{nels,
   author = {Nelson, D. and Springel, V. and Pillepich, A. and others },
   title = {The {IllustrisTNG} simulations: public data release},
   journal = {Computational Astrophysics and Cosmology},
   volume = {6},
   pages = {2},
   year = {2019}
}

@ARTICLE{nels1,
   author = {Nelson, D. and others },
   title = {First results from the {TNG50} simulation: galactic outflows driven by supernovae and black hole feedback},
   journal = {MNRAS},
   volume = {490},
   pages = { 3234–3261},
   year = {2019b}
}

@article{Pillepich_2017,
   title={First results from the IllustrisTNG simulations: the stellar mass content of groups and clusters of galaxies},
   volume={475},
   journal={MNRAS},
   author={Pillepich, Annalisa and Nelson, Dylan and Hernquist, Lars and Springel, Volker and Pakmor, Rüdiger and Torrey, Paul and Weinberger, Rainer and Genel, Shy and Naiman, Jill P and Marinacci, Federico and Vogelsberger, Mark},
   year={2017},
   pages={648–675} }

@article{Springel_2017,
   title={First results from the IllustrisTNG simulations: matter and galaxy clustering},
   volume={475},
   journal={MNRAS},
   author={Springel, Volker and Pakmor, Rüdiger and Pillepich, Annalisa and Weinberger, Rainer and Nelson, Dylan and Hernquist, Lars and Vogelsberger, Mark and Genel, Shy and Torrey, Paul and Marinacci, Federico and Naiman, Jill},
   year={2017},
   pages={676–698} }

@article{Nelson_2017,
   title={First results from the IllustrisTNG simulations: the galaxy colour bimodality},
   volume={475},
   journal={MNRAS},
   author={Nelson, Dylan and Pillepich, Annalisa and Springel, Volker and Weinberger, Rainer and Hernquist, Lars and Pakmor, Rüdiger and Genel, Shy and Torrey, Paul and Vogelsberger, Mark and Kauffmann, Guinevere and Marinacci, Federico and Naiman, Jill},
   year={2017},
   pages={624–647} }

@ARTICLE{Marinacci,
       author = {{Marinacci}, Federico and {Vogelsberger}, Mark and {Pakmor}, R{\"u}diger and {Torrey}, Paul and {Springel}, Volker and {Hernquist}, Lars and {Nelson}, Dylan and {Weinberger}, Rainer and {Pillepich}, Annalisa and {Naiman}, Jill and {Genel}, Shy},
        title = "{First results from the IllustrisTNG simulations: radio haloes and magnetic fields}",
      journal = {\mnras},
         year = 2018,
       volume = {480},
        pages = {5113-5139}
}

@ARTICLE{weinberg1995,
       author = {{Weinberg}, Martin D.},
        title = "{Production of Milky Way Structure by the Magellanic Clouds}",
      journal = {\apjl},
         year = 1995,
        month = dec,
       volume = {455},
        pages = {L31},

}

@ARTICLE{Gauthier2006,
       author = {{Gauthier}, Jean-Ren{\'e} and {Dubinski}, John and {Widrow}, Lawrence M.},
        title = "{Substructure around M31: Evolution and Effects}",
      journal = {\apj},
         year = 2006,
        month = dec,
       volume = {653},
       number = {2},
        pages = {1180-1193},

}

@ARTICLE{donghia2013,
       author = {{D'Onghia}, Elena and others},
        title = "{Self-perpetuating Spiral Arms in Disk Galaxies}",
      journal = {\apj},
         year = 2013,
        month = mar,
       volume = {766},
       number = {1},
          eid = {34},
        pages = {34},
       
}

@ARTICLE{gustaf1946,
       author = {{Str{\"o}mberg}, Gustaf},
        title = "{The Motions of the Stars Within 20 Parsecs of the Sun.}",
      journal = {\apj},
         year = 1946,
        month = jul,
       volume = {104},
        pages = {12},

}

@ARTICLE{nordstrom2004,
       author = {{Nordstr{\"o}m}, B. and others},
        title = "{The Geneva-Copenhagen survey of the Solar neighbourhood. Ages, metallicities, and kinematic properties of {\ensuremath{\sim}}14 000 F and G dwarfs}",
      journal = {\aap},
         year = 2004,
        month = may,
       volume = {418},
        pages = {989-1019},

}

@ARTICLE{sun2025,
       author = {{Sun}, Weixiang and {Shen}, Han and {Jiang}, Biwei and {Liu}, Xiaowei},
        title = "{The Age─Velocity Dispersion Relations of the Galactic Disk as Revealed by the LAMOST-Gaia Red Clump Stars}",
      journal = {\apj},
         year = 2025,
        month = feb,
       volume = {979},
       number = {2},
          eid = {103},
        pages = {103},
}

@ARTICLE{mccluskey2025,
       author = {{McCluskey}, Fiona and others},
        title = "{Stellar Velocity Dispersion versus Age: Consistency across Observations and Simulations, with the Milky Way as an Outlier}",
      journal = {arXiv e-prints},
         year = 2025,
        month = jun,
        eid = {arXiv:2506.11840},
        pages = {arXiv:2506.11840},
        
}

@ARTICLE{Naiman,
    author = {{Naiman}, Jill P. and {Pillepich}, Annalisa and {Springel}, Volker and {Ramirez-Ruiz}, Enrico and {Torrey}, Paul and {Vogelsberger}, Mark and {Pakmor}, R{\"u}diger and {Nelson}, Dylan and {Marinacci}, Federico and {Hernquist}, Lars and {Weinberger}, Rainer and {Genel}, Shy},
    title = "{First results from the IllustrisTNG simulations: a tale of two elements - chemical evolution of magnesium and europium}",
    journal = {\mnras},
    year = 2018,
    volume = {477},
    pages = {1206-1224},
}

@ARTICLE{lam2,
   author = {Wang, C. and Huang, Y. and Zhou, Y. and Zhang, H.},
   title = {Precise Masses, {Ages} of ~1.0 million {RGB} and {RC} stars observed by the {LAMOST}},
   journal = {A\&A},
   volume = {675},
   pages = {A26},
   year = {2023}
}

@ARTICLE{Liang2026,
       author = {{Liang}, Jinning and {Jiang}, Fangzhou and {Mo}, Houjun and {Benson}, Andrew and {Hopkins}, Philip F. and {Dekel}, Avishai and {Ho}, Luis C.},
        title = "{Connection between Galaxy Morphology and Dark-matter Halo Structure. II. Predicting Disk Structure from Dark-matter Halo Properties}",
      journal = {\apj},
         year = 2026,
        month = may,
       volume = {1003},
       number = {1},
          eid = {12},
        pages = {12},

}

@ARTICLE{payel,
   author = {Das, P. and Huang, Y. and Ciucă, I. and Fragkoudi, F.},
   title = {{The outer low-$\alpha$ disc of the Milky Way - I: evidence for the first pericentric passage of Sagittarius? }},
   journal = {MNRAS},
   year = {2024},
   volume ={527},
   pages = {4505–4514}
}

@ARTICLE{que2018,
       author = {{Queiroz}, A.~B.~A. and others},
       title = {{StarHorse: a Bayesian tool for determining stellar masses, ages, distances, and extinctions for field stars}},
       journal = {MNRAS},
       year = 2018,
       volume = {476},
       pages = {2556-2583},
}

@ARTICLE{que,
   author = {Queiroz, A. B. A. and Anders, F. and Chiappini, C. and Khalatyan, A. and Santiago, B. X.},
   title = {{StarHorse results for spectroscopic surveys + Gaia DR3: Chrono-chemical populations in the solar vicinity, the genuine thick disk, and young-alpha rich stars}},
   journal = {A\&A},
   volume = {673},
   pages = {A155},
   year = {2023}
}

@ARTICLE{mori,
       author = {{Mori}, A. and others},
        title = "{Metallicity distributions of halo stars: do they trace the Galactic accretion history?}",
      journal = {\aap},
         year = 2024,
        month = oct,
       volume = {690},
          eid = {A136},
        pages = {A136},

}

@ARTICLE{pillepich2023,
       author = {{Pillepich}, Annalisa and {Sotillo-Ramos}, Diego and {Ramesh}, Rahul and {Nelson}, Dylan and {Engler}, Christoph and {Rodriguez-Gomez}, Vicente and {Fournier}, Martin and {Donnari}, Martina and {Springel}, Volker and {Hernquist}, Lars},
        title = "{Milky Way and Andromeda analogues from the TNG50 simulation}",
      journal = {\mnras},
         year = 2024,
        month = dec,
       volume = {535},
       number = {2},
        pages = {1721-1762},

}

@misc{helmi1999,
      title={{Debris streams in the solar neighbourhood as relicts from the formation of the Milky Way}}, 
      author={Helmi, A. and White, S. D. M. and others},
      journal = {Nature},
      volume = {402},
      pages = {53},
      year = {1999}
}

@ARTICLE{ibata,
   author = { Ibata, R. A. and others},
   title = {{A dwarf satellite galaxy in Sagittarius.}},
   journal = {Nature},
   volume = {370},
   pages = { 194-196},
   year = {1994}
}

@article{weiss_2018,
   author={Weiss, Jake and others},
   year={2018},
   title={{A Tangle of Stellar Streams in the North Galactic Cap}},
   volume={867},
   journal={ApJ Letters},
   pages={L1} }

@article{newberg2009,
   author={Newberg, Heidi Jo and others },
   year={2009},
   title={{Discovery of a New, Polar-Orbiting Debris Stream in the Milky Way Stellar Halo}},
   volume={700},
   journal={ApJ},
   pages={L61} }

@article{majewski_2003,
   title={{A Two Micron All Sky Survey View of the Sagittarius Dwarf Galaxy. I. Morphology of the Sagittarius Core and Tidal Arms}},
   volume={599},
   journal={ApJ},
   author={Majewski, Steven R. and others},
   year={2003},
   pages={1082}}

@article{helmi2000,
   title={{ Mapping the substructure in the Galactic halo with the next generation of astrometric satellites}},
   volume={319},
   journal={MNRAS},
   author={Helmi, A. and de Zeeuw, P. T.},
   year={2000},
   pages={657}}

@ARTICLE{helmi_2006,
       author = {{Helmi}, Amina and {Navarro}, J.~F. and {Nordstr{\"o}m}, B. and {Holmberg}, J. and {Abadi}, M.~G. and {Steinmetz}, M.},
        title = "{Pieces of the puzzle: ancient substructure in the Galactic disc}",
      journal = {\mnras},
         year = 2006,
       volume = {365},
        pages = {1309-1323}}

@ARTICLE{klement_2009,
       author = {{Klement}, R. and others },
        title = "{Halo Streams in the Seventh Sloan Digital Sky Survey Data Release}",
      journal = {\apj},
         year = 2009,
       volume = {698},
        pages = {865-894}}

@ARTICLE{smith_2009,
       author = {{Smith}, M.~C. and others},
        title = "{Kinematics of SDSS subdwarfs: structure and substructure of the Milky Way halo}",
      journal = {\mnras},
         year = 2009,
       volume = {399},
        pages = {1223-1237}}

@ARTICLE{nissen_2010,
       author = {{Nissen}, P.~E. and {Schuster}, W.~J.},
        title = "{Two distinct halo populations in the solar neighborhood. Evidence from stellar abundance ratios and kinematics}",
      journal = {\aap},
         year = 2010,
       volume = {511},
        pages = {L10}}

@article{knebe,
   title={{Mapping substructures in dark matter haloes}},
   volume={357},
   journal={MNRAS},
   author={Knebe, A. and Gill, S. P. D. and Kawata, D. and Gibson, B. K.},
   year={2005},
   pages={L35}}

@article{Brown_2005,
   title={{Detection of satellite remnants in the Galactic Halo withGaia- I. The effect of the Galactic background, observational errors and sampling}},
   volume={359},
   journal={MNRAS},
   author={Brown, Anthony G. A. and others},
   year={2005},
   pages={1287} }

@article{Helmi_2005,
   title={{Pieces of the puzzle: ancient substructure in the Galactic disc: Ancient substructure in the Galactic disc}},
   volume={365},
   journal={MNRAS},
   author={Helmi, Amina and Navarro, J. F. and others},
   year={2005},
   pages={1309} }

@ARTICLE{reid2019,
       author = {{Reid}, M.~J. and others},
        title = "{Trigonometric Parallaxes of High-mass Star-forming Regions: Our View of the Milky Way}",
      journal = {\apj},
         year = 2019,
        month = nov,
       volume = {885},
       number = {2},
          eid = {131},
        pages = {131},

}

@article{gomez2010,
   title={{On the identification of merger debris in the Gaia era: Identifying merger debris in the Gaia era}},
   volume={408},
   journal={MNRAS},
   author={Gómez, Facundo A. and others},
   year={2010},
   pages={935} }

@ARTICLE{gomez2013,
       author = {{G{\'o}mez}, Facundo A. and {Minchev}, Ivan and others},
        title = "{Vertical density waves in the Milky Way disc induced by the Sagittarius dwarf galaxy}",
      journal = {\mnras},
         year = 2013,
        month = feb,
       volume = {429},
       number = {1},
        pages = {159-164},
}

@article{Font_2006,
   title="{Phase‐Space Distributions of Chemical Abundances in Milky Way–Type Galaxy Halos}",
   volume={646},
   journal={ApJ},
   author={Font, Andreea S. and Johnston, Kathryn V. and others},
   year={2006},
   pages={886} }

@article{Choi_2007,
   title="{The dynamics of tidal tails from massive satellites}",
   volume={381},
   journal={MNRAS},
   author={Choi, J. and others},
   year={2007},
   pages={987} }

@article{Morrison_2009,
   title={{Fashionably Late? Building up the Milky Way’s Inner Halo}},
   volume={694},
   journal={ApJ},
   author={Morrison, Heather L. and Helmi, Amina and others},
   year={2009},
   pages={130} }

@article{Fiorentin_2015,
   title={NEW SIGNATURES OF THE MILKY WAY FORMATION IN THE LOCAL HALO AND INNER-HALO STREAMERS IN THE ERA OF GAIA},
   volume={150},
   journal={AJ},
   author = {{Re Fiorentin}, Paola and {Lattanzi}, Mario G. and others},
   year={2015},
   pages={128} }

@article{Jean_Baptiste_2017,
   title={{On the kinematic detection of accreted streams in the Gaia era: a cautionary tale}},
   volume={604},
   journal={A\&A},
   author={Jean-Baptiste, I. and Di Matteo, P. and Haywood, M. and Gómez, A. and Montuori, M. and Combes, F. and Semelin, B.},
   year={2017},
   pages={A106} }

@article{gaia2023,
   title={{ Gaia Data Release 3}},
   volume={674},
   journal={A\&A},
   author={Gaia Collaboration., Vallenari, A. and others},
   year={2023},
   pages={A1} 
}

@misc{lamost2012,
      title={{LAMOST Spectral Survey}}, 
      author={Gang Zhao and Yongheng Zhao and Yaoquan Chu and Yipeng Jing and Licai Deng},
      year={2012},
      journal={Res. Astron. Astrophys.},
      volume={12},
      pages={723}
      }

@article{galah,
   title={{The GALAH survey: scientific motivation}},
   volume={3},
   journal={MNRAS},
   author={De Silva, G.M and others},
   year={2015},
   pages={2604}}

@article{Pagnini_2023,
   title={The distribution of globular clusters in kinematic spaces does not trace the accretion history of the host galaxy},
   volume={673},
   journal={A\&A},
   author={Pagnini, G. and Di Matteo, P. and others},
   year={2023},
   pages={A86} }

@misc{Necib2026,
      title={Galactic Amnesia: The Information Washout of the Milky Way Merger History}, 
      author={Lina Necib and Dylan Folsom and Elliot Y. Davies and Nathaniel Starkman and Andreas Thoyas},
      year={2026},
      eprint={2605.04138},
      archivePrefix={arXiv},
      primaryClass={astro-ph.GA},
      url={https://arxiv.org/abs/2605.04138}, 
}

@ARTICLE{koppelman2020,
       author = {{Koppelman}, Helmer H. and {Bos}, Roy O.~Y. and {Helmi}, Amina},
        title = "{The messy merger of a large satellite and a Milky Way-like galaxy}",
      journal = {\aap},
         year = 2020,
       volume = {642},
          eid = {L18},
        pages = {L18},
}

@ARTICLE{purcell2011,
       author = {{Purcell}, Chris W. and {Bullock}, James S. and others},
        title = "{The Sagittarius impact as an architect of spirality and outer rings in the Milky Way}",
      journal = {\nat},
         year = 2011,
        month = sep,
       volume = {477},
       number = {7364},
        pages = {301-303},

}

@article{Amarante_2022,
   title={Gastro Library. I. The Simulated Chemodynamical Properties of Several Gaia–Sausage–Enceladus-like Stellar Halos},
   volume={937},
   journal={ApJ},
   author={Amarante, João A. S. and others},
   year={2022},
   pages={12} }

@article{Khoperskov_2023b,
   title={The stellar halo in Local Group Hestia simulations: III. Chemical abundance relations for accreted and in situ stars},
   volume={677},
   journal={A\&A},
   author={Khoperskov, Sergey and others},
   year={2023},
   pages={A91} }

@article{Bailer_Jones_2021,
   title="{Estimating Distances from Parallaxes. V. Geometric and Photogeometric Distances to 1.47 Billion Stars in Gaia Early Data Release 3}",
   volume={161},
   journal={AJ},
   author={Bailer-Jones, C. A. L. and others},
   year={2021},
   pages={147} }

@INPROCEEDINGS{massa1989,
       author = {{Massa}, D. and {Savage}, B.},
        title = "{Measurements of Interstellar Extinction}",
    booktitle = {Interstellar Dust},
         year = 1989,
       editor = {{Allamandola}, Louis J. and {Tielens}, A.~G.~G.~M.},
       series = {IAU Symposium},
       volume = {135},
        month = jan,
        pages = {3},
       adsurl = {https://ui.adsabs.harvard.edu/abs/1989IAUS..135....3M}
}

@ARTICLE{cardelli1989,
       author = {{Cardelli}, Jason A. and others},
        title = "{The Relationship between Infrared, Optical, and Ultraviolet Extinction}",
      journal = {\apj},
         year = 1989,
        month = oct,
       volume = {345},
        pages = {245},
      
}

@ARTICLE{saha2010,
       author = {{Saha}, Kanak and {Tseng}, Yao-Huan and {Taam}, Ronald E.},
        title = "{The Effect of Bars and Transient Spirals on the Vertical Heating in Disk Galaxies}",
      journal = {\apj},
         year = 2010,
        month = oct,
       volume = {721},
       number = {2},
        pages = {1878-1890},

}

@ARTICLE{pavel2002,
       author = {{Kroupa}, Pavel},
        title = "{Thickening of galactic discs through clustered star formation}",
      journal = {\mnras},
         year = 2002,
        month = mar,
       volume = {330},
       number = {3},
        pages = {707-718},

}

@ARTICLE{mathis1990,
       author = {{Mathis}, John S.},
        title = "{Interstellar dust and extinction.}",
      journal = {\araa},
         year = 1990,
        month = jan,
       volume = {28},
        pages = {37-70},

}

@ARTICLE{spitzer1953,
       author = {{Spitzer}, Jr., Lyman and {Schwarzschild}, Martin},
        title = "{The Possible Influence of Interstellar Clouds on Stellar Velocities. II.}",
      journal = {\apj},
         year = 1953,
        month = jul,
       volume = {118},
        pages = {106},

}

@ARTICLE{lacey1984,
       author = {{Lacey}, C.~G.},
        title = "{The influence of massive gas clouds on stellar velocity dispersions in galactic discs}",
      journal = {\mnras},
         year = 1984,
        month = jun,
       volume = {208},
        pages = {687-707},

}

@ARTICLE{pavel2011,
       author = {{Assmann}, P. and others},
        title = "{Popping star clusters as building blocks of the Milky Way's thick disc}",
      journal = {\mnras},
         year = 2011,
        month = aug,
       volume = {415},
       number = {2},
        pages = {1280-1289},

}

@ARTICLE{medina2015,
       author = {{Martinez-Medina}, L.~A. and others},
        title = "{The Contribution of Spiral Arms to the Thick Disk Along the Hubble Sequence}",
      journal = {\apj},
         year = 2015,
        month = apr,
       volume = {802},
       number = {2},
          eid = {109},
        pages = {109},
}

@INPROCEEDINGS{roman2024,
       author = {{Schlieder}, Joshua E. and {Barclay}, Thomas and {Barnes}, Amethyst and {Bray}, Evan and {Choi}, Ami and {Cromey}, Benjamin and {Delker}, Thomas and {Finch}, Timothy and {Frater}, Eric H. and {Hill}, Robert J. and {Kruk}, Jeffrey and {Lasco}, Jeffrey and {Louie}, Dana R. and {Malhotra}, Sangeeta and {McEnery}, Julie E. and {Mosby}, Gregory and {Paine}, Jennie and {Perkins}, Jeremy S. and {Rauscher}, Bernard J. and {Rhoads}, James E. and {Rizzo}, Maxime and {Sabatke}, Derek and {Schweickart}, Rusty and {Shukis}, Diana and {Switzer}, Eric R. and {Wollack}, Edward J. and {Zellem}, Robert T. and {Zimmerman}, Neil T.},
        title = "{Survey science with the Nancy Grace Roman Space Telescope Wide Field Instrument}",
    booktitle = {Space Telescopes and Instrumentation 2024: Optical, Infrared, and Millimeter Wave},
         year = 2024,
       editor = {{Coyle}, Laura E. and {Matsuura}, Shuji and {Perrin}, Marshall D.},
       series = {Society of Photo-Optical Instrumentation Engineers (SPIE) Conference Series},
       volume = {13092},
        month = aug,
          eid = {130920S},
        pages = {130920S},
       adsurl = {https://ui.adsabs.harvard.edu/abs/2024SPIE13092E..0SS}
}

@ARTICLE{haninnen2002,
       author = {{H{\"a}nninen}, Jyrki and {Flynn}, Chris},
        title = "{Simulations of the heating of the Galactic stellar disc}",
      journal = {\mnras},
         year = 2002,
        month = dec,
       volume = {337},
       number = {2},
        pages = {731-742},

}

@ARTICLE{gallart,
       author = {{Gallart}, Carme and others},
       title = "{Uncovering the birth of the Milky Way through accurate stellar ages with Gaia}",
       journal = {Nature Astronomy},
       year = 2019,
       volume = {3},
       pages = {932-939}}

@ARTICLE{toth,
       author = {{Toth}, G. and {Ostriker}, J.~P.},
        title = {{Galactic Disks, Infall, and the Global Value of Omega}},
        journal = {ApJ},
        year = 1992,
        volume = {389},
        pages = {5}}

@ARTICLE{ostriker1989,
       author = {{Ostriker}, E.~C. and {Binney}, J.~J.},
        title = "{Warped and tilted galactic discs}",
      journal = {\mnras},
         year = 1989,
        month = apr,
       volume = {237},
        pages = {785-798},

}

@ARTICLE{quinn,
       author = {{Quinn}, P.~J. and {Hernquist}, Lars and {Fullagar}, D.~P.},
       title = {{Heating of Galactic Disks by Mergers}},
       journal =  {Astrophysical Journal},
       year = 1993,
       volume = {403},
       pages = {74}}

@article{Martig_2014,
   title={{Dissecting simulated disc galaxies – II. The age–velocity relation}},
   volume={443},
   journal={MNRAS},
   author={Martig, Marie and Minchev, Ivan and Flynn, Chris},
   year={2014},
   pages={2452–2462}}

@article{Khoperskov_2023,
   title="{The stellar halo in Local Group Hestia simulations: I. The in situ component and the effect of mergers}",
   volume={677},
   journal={A\&A},
   author={Khoperskov, Sergey and Minchev, Ivan and Libeskind, Noam and Haywood, Misha and Di Matteo, Paola and Belokurov, Vasily and Steinmetz, Matthias and Gomez, Facundo A. and Grand, Robert J. J. and Hoffman, Yehuda and Knebe, Alexander and Sorce, Jenny G. and Spaare, Martin and Tempel, Elmo and Vogelsberger, Mark},
   year={2023},
   pages={A89} }

@article{Grand_2016,
   title={Vertical disc heating in Milky Way-sized galaxies in a cosmological context},
   volume={459},
   journal={\mnras},
   author={Grand, Robert J. J. and others},
   year={2016},
   pages={199–219} }

@article{Myeong_2019,
   title={Evidence for two early accretion events that built the Milky Way stellar halo},
   volume={488},
   journal={MNRAS},
   author={Myeong, G C and others},
   year={2019},
   pages={1235–1247} }

@ARTICLE{naidu2021,
       author = {{Naidu}, Rohan P. and others},
        title = "{Reconstructing the Last Major Merger of the Milky Way with the H3 Survey}",
      journal = {\apj},
         year = 2021,
       volume = {923},
        pages = {92},
}

@ARTICLE{das2020,
       author = {{Das}, Payel and {Hawkins}, Keith and {Jofr{\'e}}, Paula},
        title = "{Ages and kinematics of chemically selected, accreted Milky Way halo stars}",
      journal = {\mnras},
         year = 2020,
       volume = {493},
        pages = {5195-5207},
}

@article{Vincenzo_2019,
   title={The Fall of a Giant. Chemical evolution of Enceladus, alias the Gaia Sausage},
   volume={487},
   journal={Monthly Notices of the Royal Astronomical Society: Letters},
   author={Vincenzo, Fiorenzo and  others},
   year={2019},
   month=may, pages={L47–L52} }

@article{Fattahi_2019,
   title={The origin of galactic metal-rich stellar halo components with highly eccentric orbits},
   volume={484},
   journal={MNRAS},
   author={Fattahi, Azadeh and others},
   year={2019},
   pages={4471–4483} }

@ARTICLE{lancaster2019,
       author = {{Lancaster}, Lachlan and {Koposov}, Sergey E. and {Belokurov}, Vasily and {Evans}, N. Wyn and {Deason}, Alis J.},
        title = "{The halo's ancient metal-rich progenitor revealed with BHB stars}",
      journal = {\mnras},
         year = 2019,
       volume = {486},
        pages = {378-389},
}

@ARTICLE{poggio2021,
       author = {{Poggio}, Eloisa and {Laporte}, Chervin F.~P. and {Johnston}, Kathryn V. and {D'Onghia}, Elena and {Drimmel}, Ronald and {Grion Filho}, Douglas},
        title = "{Measuring the vertical response of the Galactic disc to an infalling satellite}",
      journal = {\mnras},
         year = 2021,
        month = nov,
       volume = {508},
       number = {1},
        pages = {541-559},

}

@ARTICLE{sch2019,
       author = {{Sch{\"o}nrich}, Ralph and {McMillan}, Paul and {Eyer}, Laurent},
        title = "{Distances and parallax bias in Gaia DR2}",
      journal = {\mnras},
         year = 2019,
        month = aug,
       volume = {487},
       number = {3},
        pages = {3568-3580},
}

@ARTICLE{luri2018,
       author = {{Luri}, X. and {Brown}, A.~G.~A. and {Sarro}, L.~M. and {Arenou}, F. and {Bailer-Jones}, C.~A.~L. and {Castro-Ginard}, A. and {de Bruijne}, J. and {Prusti}, T. and {Babusiaux}, C. and {Delgado}, H.~E.},
        title = "{Gaia Data Release 2. Using Gaia parallaxes}",
      journal = {\aap},
         year = 2018,
        month = aug,
       volume = {616},
          eid = {A9},
        pages = {A9},
}

@article{Ruiz_Lara_2022,
   title={Substructure in the stellar halo near the Sun: II. Characterisation of independent structures},
   volume={665},
   journal={A\&A},
   author={Ruiz-Lara, T. and Matsuno, T. and Lövdal, S. S. and Helmi, A. and Dodd, E. and Koppelman, H. H.},
   year={2022},
   pages={A58} }

@article{Chandra_2023,
   title={Distant Echoes of the Milky Way’s Last Major Merger},
   volume={951},  
   journal={ApJ},
   publisher={American Astronomical Society},
   author={Chandra, Vedant and others},
   year={2023},
   pages={26} }

@ARTICLE{haywood2018,
       author = {{Haywood}, M. and others},
        title = "{In Disguise or Out of Reach: First Clues about In Situ and Accreted Stars in the Stellar Halo of the Milky Way from Gaia DR2}",
      journal = {\apj},
         year = 2018,
       volume = {863},
        pages = {113},
    
}

@ARTICLE{hawthorn2021,
       author = {{Bland-Hawthorn}, Joss and {Tepper-Garc{\'\i}a}, Thor},
        title = "{Galactic seismology: the evolving 'phase spiral' after the Sagittarius dwarf impact}",
      journal = {\mnras},
         year = 2021,
        month = jul,
       volume = {504},
       number = {3},
        pages = {3168-3186},

}

@ARTICLE{laporte2018,
       author = {{Laporte}, Chervin F.~P. and others},
        title = "{The influence of Sagittarius and the Large Magellanic Cloud on the stellar disc of the Milky Way Galaxy}",
      journal = {\mnras},
         year = 2018,
        month = nov,
       volume = {481},
       number = {1},
        pages = {286-306},

}

@ARTICLE{widrow2014,
       author = {{Widrow}, Lawrence M. and others},
        title = "{Bending and breathing modes of the Galactic disc}",
      journal = {\mnras},
         year = 2014,
        month = may,
       volume = {440},
       number = {3},
        pages = {1971-1981},
          
}

@misc{thulasidharan2024,
      title={The Age-Thickness Relation of the Milky Way Disk: A Tracer of Galactic Merging History}, 
      author={Lekshmi Thulasidharan and Elena D'Onghia and Robert Benjamin and Ronald Drimmel and Eloisa Poggio and Anna Queiroz},
      year={2024},
      eprint={2412.12304},
      archivePrefix={arXiv},
      primaryClass={astro-ph.GA},
      url={https://arxiv.org/abs/2412.12304}, 
}

@ARTICLE{nataf2024,
       author = {{Nataf}, David M. and {Schlaufman}, Kevin C. and {Reggiani}, Henrique and {Hahn}, Isabel},
        title = "{Accurate, Precise, and Physically Self-consistent Ages and Metallicities for 400,000 Solar Neighborhood Subgiant Branch Stars}",
      journal = {\apj},
  
         year = 2024,
        month = nov,
       volume = {976},
       number = {1},
          eid = {87},
        pages = {87},

}

@ARTICLE{chaplin2013,
       author = {{Chaplin}, William J. and {Miglio}, Andrea},
        title = "{Asteroseismology of Solar-Type and Red-Giant Stars}",
      journal = {\araa},
         year = 2013,
        month = aug,
       volume = {51},
       number = {1},
        pages = {353-392},
       
}

@ARTICLE{fred2004,
       author = {{Pont}, Fr{\'e}d{\'e}ric and {Eyer}, Laurent},
        title = "{Isochrone ages for field dwarfs: method and application to the age-metallicity relation}",
      journal = {\mnras},
         year = 2004,
        month = jun,
       volume = {351},
       number = {2},
        pages = {487-504},
}

@ARTICLE{jorgensen2005,
       author = {{J{\o}rgensen}, B.~R. and {Lindegren}, L.},
        title = "{Determination of stellar ages from isochrones: Bayesian estimation versus isochrone fitting}",
      journal = {\aap},
         year = 2005,
        month = jun,
       volume = {436},
       number = {1},
        pages = {127-143},
}

@ARTICLE{Feuillet2018,
       author = {{Feuillet}, Diane K. and {Bovy}, Jo and {Holtzman}, Jon and {Weinberg}, David H. and {Garc{\'\i}a-Hern{\'a}ndez}, D. and {Hearty}, Fred R. and {Majewski}, Steven R. and {Roman-Lopes}, Alexandre and {Rybizki}, Jan and {Zamora}, Olga},
        title = "{Age-resolved chemistry of red giants in the solar neighbourhood}",
      journal = {\mnras},
         year = 2018,
        month = jun,
       volume = {477},
       number = {2},
        pages = {2326-2348},

}

@ARTICLE{Almannaei2026,
       author = {{Almannaei}, Aisha S. and others},
        title = "{Towards Galactic archaeology with inferred ages of giant stars from Gaia spectra}",
      journal = {\mnras},
         year = 2026,
        month = feb,
       volume = {546},
       number = {1},
          eid = {staf2252},
        pages = {staf2252},
}

@ARTICLE{lsst2019,
       author = {{Ivezi{\'c}}, {\v{Z}}eljko and {Kahn}, Steven M. and others},
      journal = {\apj},
         year = 2019,
       volume = {873},
        pages = {111},

}

@ARTICLE{roman,
       author = {{Han}, Jiwon Jesse and {Chiti}, Anirudh and others},
        title = "{A Path to an All-Sky Survey with Roman}",
      journal = {arXiv e-prints},
         year = 2026,
        month = feb,
        pages = {arXiv:2602.21280},

}

@ARTICLE{LianLuo2024,
       author = {{Lian}, Jianhui and {Luo}, Li},
        title = "{The Thickness of Galaxy Disks from z = 5 to 0 Probed by JWST}",
      journal = {\apjl},
         year = 2024,
        month = jan,
       volume = {960},
       number = {2},
          eid = {L10},
        pages = {L10},
          doi = {10.3847/2041-8213/ad1492},
archivePrefix = {arXiv},
       eprint = {2312.07070},
 primaryClass = {astro-ph.GA},
       adsurl = {https://ui.adsabs.harvard.edu/abs/2024ApJ...960L..10L}
}

@ARTICLE{Tsukui2025,
       author = {{Tsukui}, Takafumi and {Wisnioski}, Emily and {Bland-Hawthorn}, Joss and {Freeman}, Ken},
        title = "{The emergence of galactic thin and thick discs across cosmic history}",
      journal = {\mnras},
         year = 2025,
        month = jul,
       volume = {540},
       number = {4},
        pages = {3493-3522},
          doi = {10.1093/mnras/staf604},
archivePrefix = {arXiv},
       eprint = {2409.15909},
 primaryClass = {astro-ph.GA},
       adsurl = {https://ui.adsabs.harvard.edu/abs/2025MNRAS.540.3493T}
}

\appendix
\setcounter{figure}{0}
\renewcommand{\thefigure}{A\arabic{figure}}
\renewcommand{\thesubsection}{A.\arabic{subsection}}
\setcounter{subsection}{0}
\subsection{Robustness of the merger signature to accreted populations}

\begin{figure}[!htp]
    \centering 
    \includegraphics[width=\columnwidth]{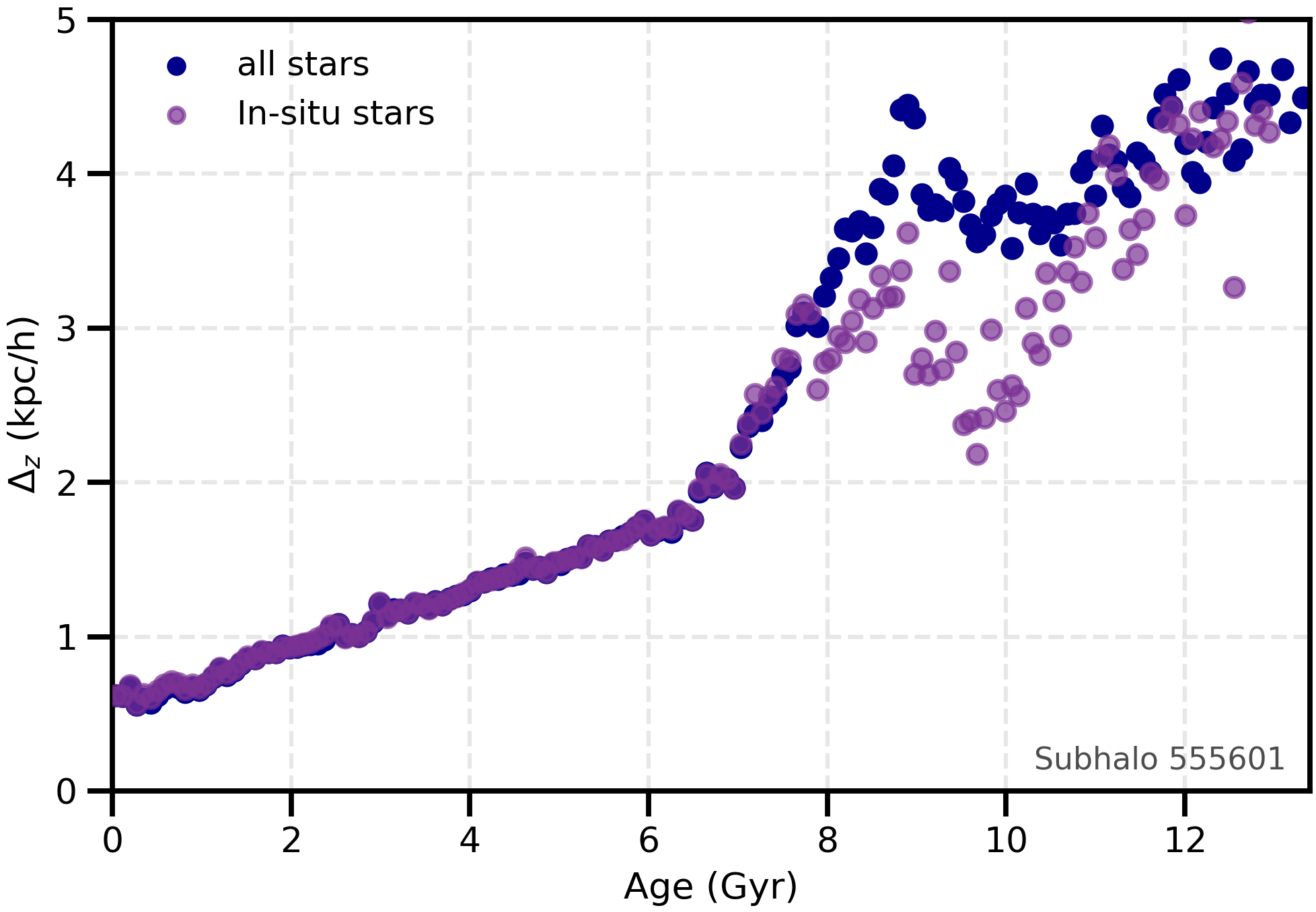}
    \caption{Age--$\Delta_z$ relation for Subhalo 555601, comparing the fiducial sample of all stars with the subset of stars flagged as in-situ by the \texttt{InSitu} flag in the TNG50 snapshot. The merger-associated enhancement in $\Delta_z$ remains visible in the in-situ-selected sample, indicating that the feature is not solely driven by accreted stellar debris.}
    \label{fig:A1}
\end{figure}

To ensure that the merger signature is not solely a contamination from accreted populations, as an additional check, we repeated the analysis for galaxy 555601, using only star particles marked as in-situ by the available ‘InSitu’ flag in the present-day snapshot file. The resulting age--$\Delta_z$ relation is compared with the relation obtained using all stars in Figure \ref{fig:A1}. This comparison shows that the ex-situ stars do contribute to the enhanced thickness during the merger epoch. However, the main merger-associated increase in thickness remains clearly visible in the in-situ-only sample as well. The feature is therefore not produced solely by stars identified as accreted debris.

\subsection{Evolution of the Age--$\Delta_z$ and Age--$\sigma_z$ relations}
\label{sec:thickness_ev}
To trace the age-thickness relation evolution as a function of time for the snapshots before the merger interaction, we selected stars belonging to the main progenitor subhalo at each snapshot. Because the galaxy grows substantially with time, we did not impose the present-day physical radial range on all earlier snapshots. Instead, at each snapshot, we fitted the stellar surface-density profile with an exponential disk and measured the scale length, $R_d (t)$. We then computed the age--$\Delta_z$ relation for stars in the radial range $R_d(t)<R<3R_d(t)$. 
\begin{figure}[!htp]
    \centering 
    \includegraphics[width=\columnwidth]{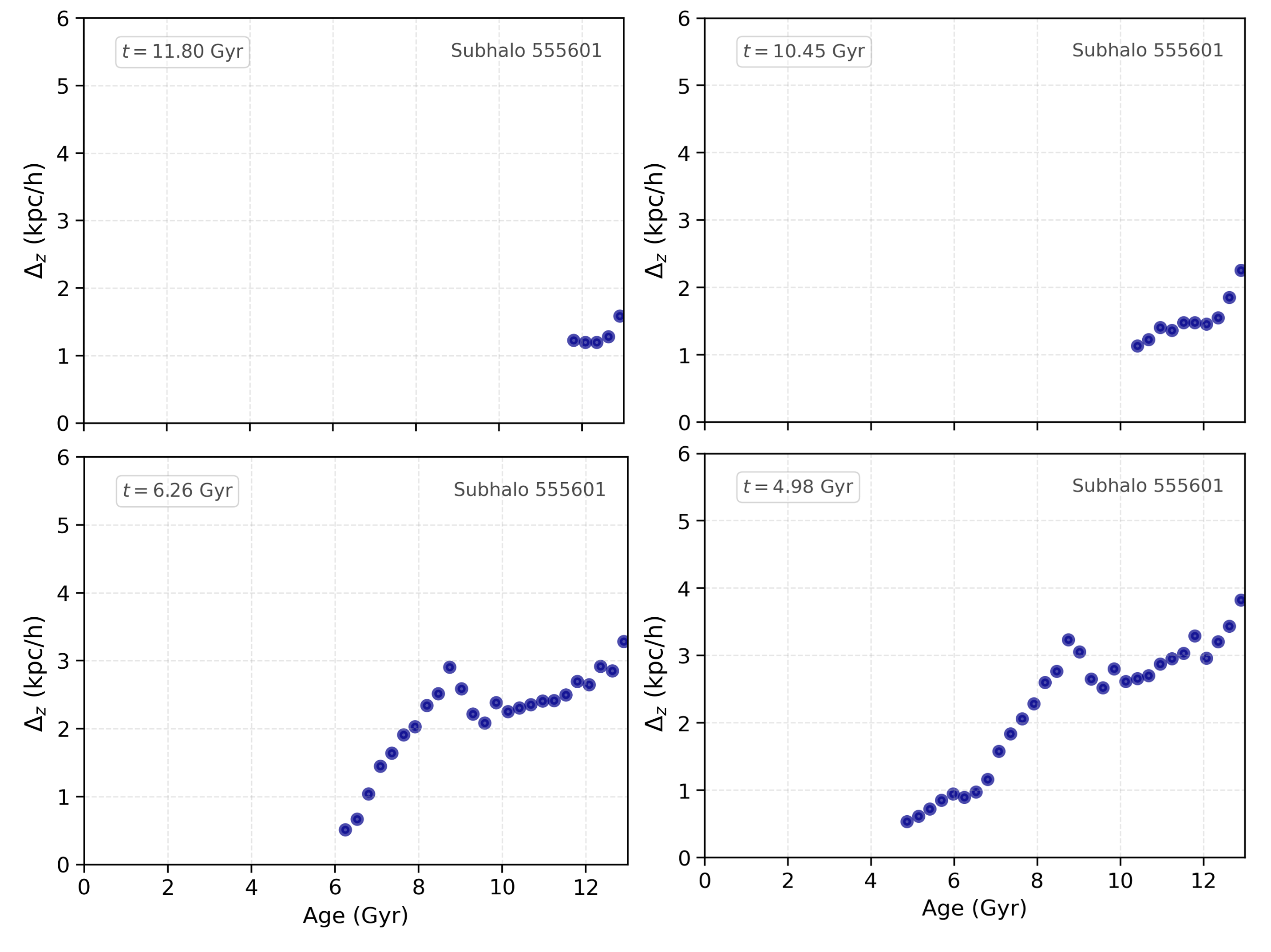}
    \caption{Evolution of the age--$\Delta_z$ relation for Subhalo~555601. The first two panels show pre-interaction snapshots at lookback times of $11.80$ and $10.45$ Gyr, while the latter two panels show post-interaction snapshots at lookback times of $6.26$ and $4.98$ Gyr. Stellar ages are shown as present-day ages for consistency.}
    \label{fig:A4}
\end{figure}
\begin{figure}[!htp]
    \centering 
    \includegraphics[width=\columnwidth]{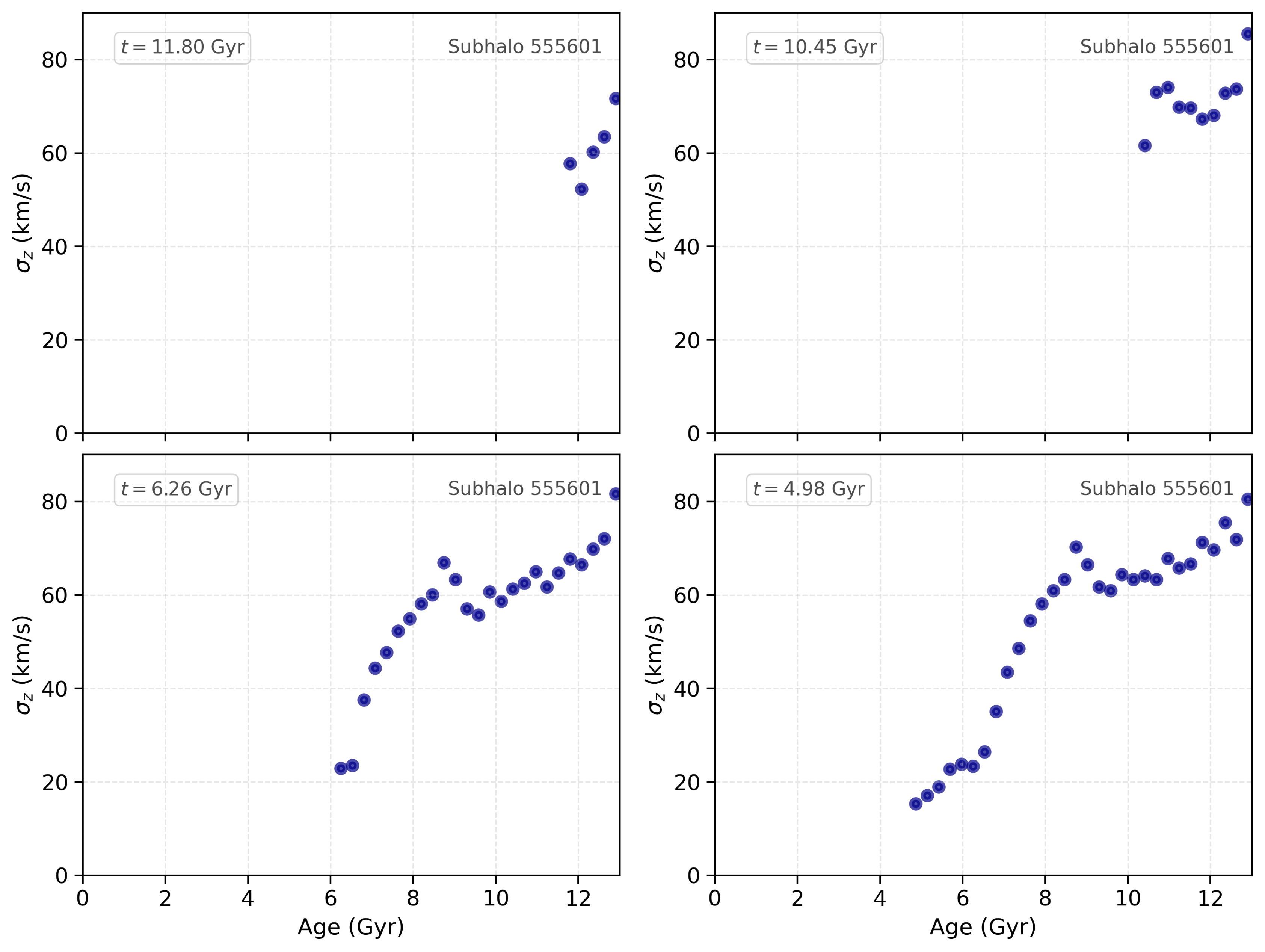}
    \caption{Same as Figure \ref{fig:A4} but for age--$\sigma_z$ relation.}
    \label{fig:A5}
\end{figure}

Figure \ref{fig:A4} shows the evolution of the age--$\Delta_z$ relation for Subhalo~555601 at four snapshots, while Figure \ref{fig:A5} shows the corresponding evolution of the age--$\sigma_z$ relation. In both panels, the first two columns correspond to pre-interaction snapshots at $t=11.80$ and $10.45$ Gyr, while the latter two columns show post-interaction snapshots at $t=6.26$ and $4.98$ Gyr.

In the pre-interaction snapshots, the age--$\Delta_z$ relation follows a monotonic trend among the stellar populations present at the respective lookback time. This behavior is consistent with the expectation that stars formed earlier were born in a dynamically hotter disk than stars formed later. The corresponding age--$\sigma_z$ relation shows the same qualitative behavior, indicating that the older populations were already kinematically hot before the merger. However, the amplitude of $\Delta_z$ measured at the pre-interaction snapshots is smaller than that measured in the present-day snapshot for the same present-day stellar age range (Figure \ref{fig:fig2}). This is expected because the early galaxy is compact, with a much smaller scale length and a centrally dense stellar distribution, so the stronger vertical restoring force can keep even dynamically hot populations spatially confined, for a given vertical velocity dispersion. As the disk evolves, its scale length also grows, especially after the merger event, so the radial range sampled by $R_d(t)<R<3R_d(t)$ and the local vertical restoring force change with time. As the disk grows outward, the sampled radial range extends into lower-density regions where the local vertical restoring force is weaker, allowing stellar populations with similar vertical motions to occupy a larger scale height.

The post-interaction snapshots show a stronger vertical separation between stars formed before, during, and after the interaction. In particular, the age--$\Delta_z$ relation becomes more vertically extended and develops the step-like feature associated with the merger epoch. A similar trend is also visible in the age--$\sigma_z$ relation. We therefore interpret the pre-interaction trends as the underlying upside-down vertical structure of the disk, and the step-like signature and enhanced thickness as the response of this already kinematically hot stellar disk to the interaction.

A smooth monotonic present-day age--$\Delta_z$ relation alone cannot uniquely distinguish between stars that were born dynamically hot in an upside-down assembly scenario and stars that were born colder but subsequently heated over time. However, the localized step-like feature associated with the merger epoch provides an additional diagnostic. It marks a transition between stars formed before or during a dynamically disturbed phase and stars formed after the gas disk settled into a colder, thinner configuration. This supports an upside-down vertical assembly picture, with the merger imprint superposed on the underlying age-dependent vertical structure.

This analysis shows that the interaction signatures identified in the present-day relation remain physically meaningful, especially where localized steps or bumps coincide with merger epochs. However, the absolute amplitude of $\Delta_z$ should not be interpreted as a direct measure of merger-driven heating alone. Instead, it reflects the combined effects of birth thickness, merger-driven and secular heating, radial redistribution, and the evolving vertical restoring force that maps vertical motions into a physical scale height.

\subsection{Testing for non-disk contamination in Subhalo 555601}
\label{app:555601_morphology}

To check whether the large values of $\Delta_z$ in Subhalo 555601 reflect a disk-like stellar component rather than a purely non-disk or spheroidal component, we examined the morphology and kinematics of the same age-selected populations. We divided the stars in the radial range $R_d < R < 3R_d$ into three age intervals: ages younger than $6.6$ Gyr, corresponding to the geometrically thinner disk formed after the merger; ages between $6.6$ and $10.3$ Gyr, corresponding to the merger epoch; and ages older than $10.3$ Gyr, corresponding to stars formed before the merger epoch.

\begin{figure}[!htp]
    \centering 
    \includegraphics[width=\columnwidth]{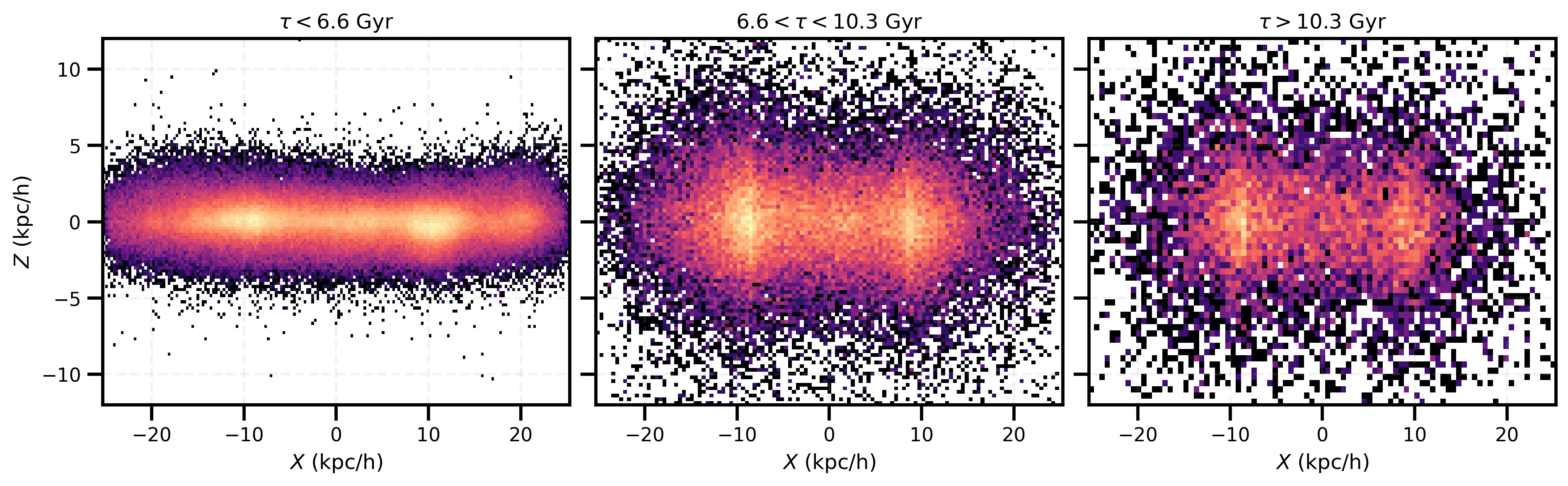}
    \caption{Present-day $X-Z$ density distributions of stars in Subhalo 555601 for three stellar age intervals, measured in the radial range $R_d<R<3R_d$.}
    \label{fig:A2}
\end{figure}

For each population, we show the present-day $X-Z$ density distribution in Figure \ref{fig:A2}. The merger-age and pre-merger populations are more vertically extended than the younger post-merger population, as expected. However, their distributions remain flattened: in all three age intervals, the horizontal extent is substantially larger than the vertical extent. To quantify this, we computed the ratio $\mathrm{std}(z)/\mathrm{std}(x)$, which provides a simple measure of the vertical extent relative to the horizontal extent in this projection. We find $\mathrm{std}(z)/\mathrm{std}(x)=0.104$, $0.31$, and $0.39$ for the $<6.6$ Gyr, $6.6-10.3$ Gyr, and $>10.3$ Gyr populations, respectively. A value close to 1 would indicate comparable vertical and horizontal extents pointing to a spheroidal distribution, while all three populations have values less than 0.5. Thus although the older populations are thicker, they remain closer to a flattened distribution.

\begin{figure}[!htp]
    \centering 
    \includegraphics[width=\columnwidth]{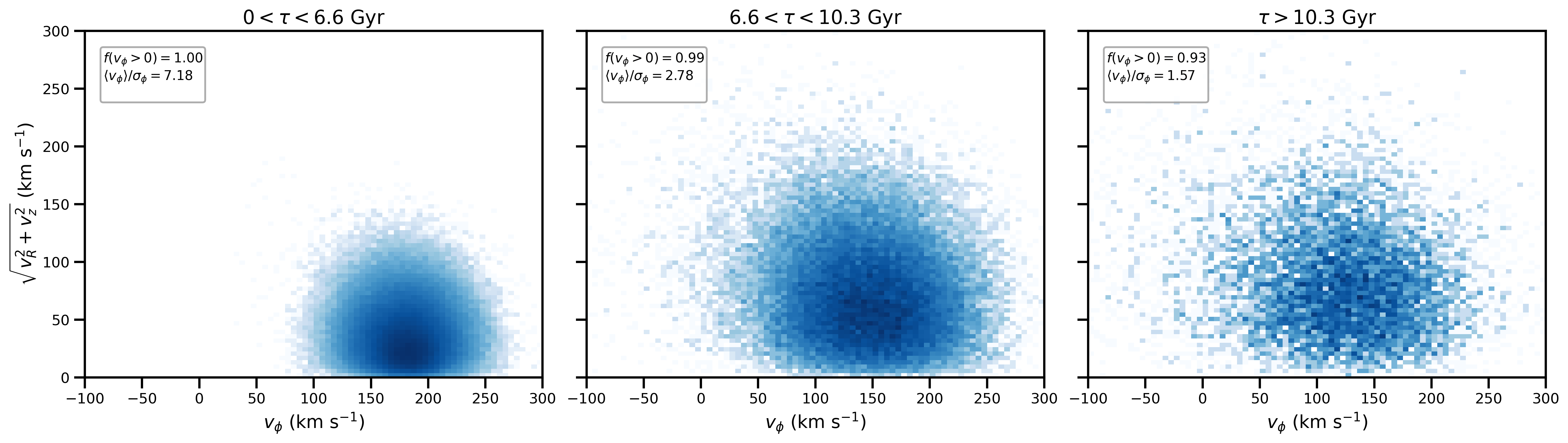}
    \caption{Kinematic distributions of the same age-selected stellar populations as in Figure \ref{fig:A2}, shown in the $v_\phi$--$\sqrt{v_R^2+v_z^2}$ plane. The younger post-merger population has a larger characteristic $v_\phi$ and lower non-azimuthal velocity amplitude, consistent with a colder disk component. The merger-age and pre-merger populations are dynamically hotter, but they retain net prograde rotation.}
    \label{fig:A3}
\end{figure}
As an additional kinematic diagnostic, we examined the distribution of $v_\phi$ versus $\sqrt{v_R^2+v_z^2}$, for the same three age intervals. As shown in Figure \ref{fig:A3}, the stars born after the merger have a larger characteristic $v_\phi$ and a smaller range of $\sqrt{v_R^2+v_z^2}$, as expected for a colder disk component. The merger-age and pre-merger populations have a lower characteristic $v_\phi$ and a broader range of $\sqrt{v_R^2+v_z^2}$, indicating that they are dynamically hotter. However, both populations still show net prograde rotation. This is quantified by the fraction of stars in the prograde orbits, $(f(v_\phi>0))$, which remains high across all three age intervals: 1.00, 0.99, and 0.93 from the youngest to oldest populations. At the same time, $\langle v_\phi\rangle/\sigma_\phi$ decreases from 7.18 to 2.78 and 1.57, indicating that the older populations are progressively less supported by ordered rotation.

The flattened $X-Z$ morphology shown in Figure \ref{fig:A2}, together with the net prograde rotation in Figure \ref{fig:A3}, indicates that the large thickness values in Subhalo 555601 arise from dynamically hot but still disk-like stellar populations. Thus, the merger-epoch and pre-merger epoch stellar populations are better described as a geometrically thick disk component than as a spheroidal stellar halo.

\end{document}